\newif\iffigs\figstrue
\documentclass[12pt]{article}
\usepackage{latexsym,amssymb}
\usepackage{latexsym,amssymb}
\iffigs
      \input{epsf}
\else
\fi
\def\IC{\relax\,\hbox{$\inbar\kern-.3em{\rm C}$}}
\def\IG{\relax\,\hbox{$\inbar\kern-.3em{\rm G}$}}
\def\IB{\relax{\rm I\kern-.18em B}}
\def\ID{\relax{\rm I\kern-.18em D}}
\def\IL{\relax{\rm I\kern-.18em L}}
\def\IF{\relax{\rm I\kern-.18em F}}
\def\IH{\relax{\rm I\kern-.18em H}}
\def\II{\relax{\rm I\kern-.17em I}}
\def\IN{\relax{\rm I\kern-.18em N}}
\def\IP{\relax{\rm I\kern-.18em P}}
\def\IQ{\relax\,\hbox{$\inbar\kern-.3em{\rm Q}$}}
\def\bfzero{\relax\,\hbox{$\inbar\kern-.3em{\rm 0}$}}
\def\IK{\relax{\rm I\kern-.18em K}}
\def\IG{\relax\,\hbox{$\inbar\kern-.3em{\rm G}$}}
 \font\cmss=cmss10 \font\cmsss=cmss10 at 7pt
\def\IR{\relax{\rm I\kern-.18em R}}
\def\ZZ{\relax\ifmmode\mathchoice
{\hbox{\cmss Z\kern-.4em Z}}{\hbox{\cmss Z\kern-.4em Z}}
{\lower.9pt\hbox{\cmsss Z\kern-.4em Z}}
{\lower1.2pt\hbox{\cmsss Z\kern-.4em Z}}\else{\cmss Z\kern-.4em
Z}\fi}
\def\bfone{\relax{\rm 1\kern-.35em 1}}

\def\inbar{\vrule height1.5ex width.4pt depth0pt}
\def\bfzero{\relax{\rm I\kern-.18em 0}}
\def\bfone{\relax{\rm 1\kern-.35em 1}}

\newcommand{\ft}[2]{{\textstyle\frac{#1}{#2}}}

\def\1bar{1\hskip -.275cm -}
\def\2bar{2\hskip -.275cm -}
\def\3bar{3\hskip -.275cm -}

\newsavebox{\uuunit}
\sbox{\uuunit}
        {\setlength{\unitlength}{0.825em}
          \begin{picture}(0.6,0.7)
                 \thinlines
                 \put(0,0){\line(1,0){0.5}}
                 \put(0.15,0){\line(0,1){0.7}}
                 \put(0.35,0){\line(0,1){0.8}}
                \multiput(0.3,0.8)(-0.04,-0.02){10}{\rule{0.5pt}{0.5pt}}
          \end {picture}}

\makeatletter \@addtoreset{equation}{section} \makeatother

\newcommand{\be}{\begin{equation}}
\newcommand{\ee}{\end{equation}}
\newcommand{\ba}{\begin{eqnarray}}
\newcommand{\ea}{\end{eqnarray}}
\def\bfone{\relax{\rm 1\kern-.35em 1}}

\def\bfone{\relax{\rm 1\kern-.35em 1}}
\font\cmss=cmss10 \font\cmsss=cmss10 at 7pt
\begin{document}
\begin{titlepage}
\vskip 1.5cm
\begin{center}
{\LARGE \bf  Exact solutions for Bianchi type cosmological metrics,
 Weyl orbits of $E_{8(8)}$ subalgebras and $p$--branes $^ \dagger $}\nonumber \\
\vfill {\large
 P. Fr\'e$^1$, K. Rulik$^1$, M. Trigiante$^2$} \nonumber \\
\vfill {
$^1$ Dipartimento di Fisica Teorica, Universit\'a di Torino,
$\&$ INFN -
Sezione di Torino\nonumber \\
via P. Giuria 1, I-10125 Torino, Italy  }\nonumber \\
\vskip 0.3cm
{
$^2$ Dipartimento di Fisica Politecnico di Torino, C.so Duca degli Abruzzi,
24,
I-10129 Torino }
\end{center}
\vfill
\begin{abstract}
{In this paper we pursue further a programme initiated in a
previous work and aimed at the construction, classification and
property investigation of time dependent solutions of supergravity
(superstring backgrounds) through a systematic exploitation of
$\mathrm{U}$--duality hidden symmetries. This is done by first
reducing to $D=3$ where the bosonic part of the theory becomes a
sigma model on $E_{8(8)}/SO(16)$, solving the equations through an
algorithm that produces general integrals for any chosen regular
subalgebra $\mathbf{G}_r \subset E_{8(8)}$ and then oxiding back
to $D=10$. Different oxidations and hence different physical
interpretations of the same solutions are associated with
different embeddings of $\mathbf{G}_r$. We show how such
embeddings constitute orbits under the Weyl group and we study the
orbit space. This is relevant to associate candidate superstring
cosmological backgrounds to space $Dp$--brane configurations that
admit microscopic descriptions. In particular in this paper we
show that there is just one Weyl orbit of $A_r$ subalgebras for $r
< 6$. The orbit of the previously found $A_2$ solutions, together
with space--brane representatives contains a pure metric
representative that corresponds to homogeneous Bianchi type 2A
cosmologies in $D=4$ based on the Heisenberg algebra. As a
byproduct of our methods we obtain new exact solutions for such
cosmologies with and without matter. We present a thorough
investigation of their properties.}
\end{abstract}
\vspace{2mm} \vfill
\hrule width 3.cm {\footnotesize
$^ \dagger $
This work is supported in part by
the European Union RTN contracts
HPRN-CT-2000-00122 and HPRN-CT-2000-00131.}
\end{titlepage}
\newpage
\section{Introduction}
Cosmological solutions of supergravity or, more generally, time
dependent superstring backgrounds have attracted a great deal of
interest in the recent years
\cite{Kachru:2003sx,Kachru:2003aw,Fre:2002pd,Gutperle:2002ai,iva,cornalba,Papadopoulos:2002bg,que,GV,craps,ban,setu,cope,mart,Sen:2002vv,Sen:2002nu},
both because of the stimulus provided by new data in observational
cosmology \cite{experiment}, which seem to imply a small but non
vanishing cosmological constant a flat geometry of the universe
and confirm inflation \cite{linde90}, and also for intrinsic
conceptual reasons inherent to a continuously sought for deeper
understanding of the internal structure of the theory.
\par
Relying on the motivations outlined above, in a recent paper \cite{piervoiastatia} we
have addressed the question of establishing a general classification
of all time dependent backgrounds of ten dimensional superstring,
providing also an algorithm capable of constructing explicit analytic
solutions of supergravity field equations. Our method is based on a systematic use of $\mathrm{U}$--duality
and exploits the algebraic structure of hidden symmetries. Indeed we
heavily relied on the observation that looking for solutions that
depend only on one--parameter is tantamount as dimensionally reducing
the theory to $1+0$ dimensions. It is also equivalent to looking at
one--parameter (time) solutions of the theory  reduced
to any dimension $d$ in the interval
\begin{equation}
  1+0 \le d \le 1+9
\label{intervallo}
\end{equation}
Preferred choice for us was $d=1+2$ since, there, all bosonic
degrees of freedom correspond to scalar fields and supergravity is
replaced by the sigma model $\mathrm{E_{8(8)}}/\mathrm{SO(16)}$.
Using the solvable Lie algebra representation of non--compact
cosets we were able to rewrite the sigma model field equations in
\textit{Nomizu operator form} and construct an algorithm that
allows:
\begin{description}
  \item[a] to obtain analytic exact solutions of the $D=3$ equations in a systematic way,
  \item[b] to \textit{oxide} back such solutions to $D=10$
  supergravity backgrounds by following precise algebraic oxidation
  rules based on a one-to-one correspondence between the $D=10$ fields
  and the $8$ Cartan generators plus $120$ positive roots of $E_{8(8)}$.
\end{description}
In particular in \cite{piervoiastatia} we proved that the possible
$D=3$ solutions are classified by regularly embedded subalgebras $
\mathbf{G}_r \subset \mathrm{E_{8}}$ of rank $r \leq 8$ and that
their ten--dimensional physical interpretation (oxidation) depends
on the classification of the different embeddings $ \mathbf{G}_r
\hookrightarrow \mathrm{E_8}$.\footnote{We recall that the linear
span with real coefficients of the Cartan-Weyl generators
$\{\mathcal{H}_i, \mathrm{E}_\alpha, \mathrm{E}_{-\alpha}\}$ of
any simple Lie algebra ${\bf G}_r$ corresponds to the maximally
non-compact section ${\bf G}_{r(r)}$ and that the exponential of
its Borel subalgebra $\{\mathcal{H}_i, \mathrm{E}_\alpha \}$
describes the maximally non-compact coset
$\mathrm{G}_{r(r)}/\mathrm{H}_r$, where $\mathrm{H}_r\subset
\mathrm{G}_{r(r)}$is the maximal compact subgroup. Therefore,
regular embeddings of Lie algebras canonically embed Borel
subalgebras into Borel subalgebras and hence maximally non-compact
cosets into maximally non-compact cosets.} We gave some
preliminary examples of explicit solutions based on the simplest
choice $\mathbf{G}_r=\mathrm{A_2}$. It also turned out that these
solutions provide a smooth and exact realization of the bouncing
phenomenon on Weyl chamber walls envisaged by the cosmological
billiards of Damour et al.
\cite{bill99,dualiza2,Henneaux:2003kk,deBuyl:2003za,Damour:2002tc,Damour:pq,Damour:2001sa,Damour:2000hv,Damour:2000th,Damour:2000wm,Demaret:sg}.
We also showed how this physical phenomenon was triggered by the
presence of extended objects possibly interpretable, at the
microscopic level, as space $Dp$--branes.
\par
In the present paper we address and answer the important related
question: \textit{How many different orbits of cosmological
solutions are there under the full $U$--duality group?} It is
interesting \textit{per se} to know that apparently completely
different classical configurations are related by duality
transformations but an exhaustive organization of
$\mathrm{U}$--orbits has a specific intrinsic value in view of the
following consideration that was already systematically exploited
in the case of supersymmetric black--holes
\cite{bertolotrigiaseries}. Suppose that a certain configuration
of supergravity fields is in the same $\mathrm{U}$--orbit as
another one which can be solely described in terms of
Ramond--Ramond $p$--forms and therefore of $D$--branes. It follows
that a microscopic stringy description is available for all
configurations sitting in that orbit. Any stringy calculation
which might be of interest can be performed for the orbit
representative where the $D$--brane description is available and
the results can be exported, via $\mathrm{U}$--duality, to all the
other cases for which a  $D$--brane boundary state cannot be
constructed. This way of reasoning was for instance used in
\cite{bertolotrigiaseries} to calculate the statistical entropy of
black--holes and verify that it coincides with the area of the
event horizon. After constructing a generating solution of
$\mathcal{N}=8$ black--holes, that depends on $5$--parameters, and
deriving  its orbit under the $\mathrm{U}$--duality group
$\mathrm{E_{7(7)}}$ it was possible to identify the $Dp$--brane
systems that corresponds to specific representatives of the orbit.
There one could perform the microscopic calculation of the
statistical entropy.
\par
A similar situation can now be realized in the context of
cosmological, rather than black--hole solutions. By a systematic
study of $\mathrm{U}$--orbits we can connect time--dependent
backgrounds of Neveu-Schwarz character, for instance purely
gravitational solutions, which might  have special physical
interest but do not admit $D$--brane descriptions to others that
are realized by $D$--brane systems, in particular by
space--branes.
\par
A  noticeable example of such relations, which has by itself
a special interest, will be illustrated in the present paper.
\par
Approximately ninety years ago, just after General Relativity was
introduced,  Bianchi classified $D=4$ cosmological metrics that are homogeneous \cite{Bianchiorinal}, namely that admit
a three parameter  group of isometries acting transitively on constant time
slices. Bianchi metrics are of the following form
\begin{eqnarray}
  ds^2&=& -A(t) \, dt^2 + h_{ij}(t) \, \Omega^i \, \otimes \,
  \Omega^{j}
  \label{Bianchimet}
\end{eqnarray}
where
\begin{equation}
  h_{ij}(t) \mbox{ is a symmetric, time dependent $3 \times 3$-matrix }
\label{Bianchimet1}
\end{equation}
and where $\Omega^i$ are the Maurer Cartan $1$--forms on a
three--dimensional group manifold:
\begin{eqnarray}
  d\Omega^i & = & t^{i}_{\phantom{i}{jk}} \Omega^j \, \wedge \,
  \Omega^{k} \quad ; \quad \left( \mbox{Maurer Cartan eq.s} \right) \nonumber\\
      t^{i}_{\phantom{i}{jk}} &=& \mbox{structure constants of a three parameter Lie
      algebra}
\label{Bianchimet2}
\end{eqnarray}
Bianchi classification of homogeneous cosmologies is a
classification of all three--dimensional algebras $\mathcal{G}_3$,
identified by the structure constants $ t^{i}_{\phantom{i}{jk}}$,
a classification that includes all non--semisimple and solvable
algebras. Once the algebra $\mathcal{G}_3$ is chosen, one is still
confronted with the task of solving Einstein equations for the
generalized scale factors $A(t)$ and $h_{ij}(t)$ in presence of
suitable matter encoded into a suitable stress--energy tensor.
Here the list of exact analytic solutions is not abundant
\cite{Bianchiorinal} since the differential equations to be solved
have for most algebras, except the abelian $\mathbb{R}^3$, a
rather formidable non linear structure that impedes to obtain
general or particular integrals.
\par
As a by--product of our orbit analysis we will present some exact
analytic solutions of Einstein equations for homogeneous cosmologies
that, up to our knowledge, were so far unknown in the literature.
\par
Specifically we will prove the following. Among the various three
parameter algebras one has the \textit{Heisenberg algebra} defined by
the following Maurer--Cartan equations:
\begin{eqnarray}
 && \nonumber d\Omega^1   +   \frac{\varpi}{2} \, \Omega^2\wedge\Omega^3 = 0 \\
 &&  d\Omega^2 = 0  \\
 &&d\Omega^3 = 0
\label{MaurerCartan}
\end{eqnarray}
or, alternatively, by the commutation relations of the dual generators $\Omega^i(T_j)=\delta^i_j$:
\begin{eqnarray}
Heis &:& \nonumber\\
 \left [ T_i , T_j \right ] & = & t_{ij}^k \, T_k \quad ; \quad
t_{23}^1 = - \frac{\varpi}{4}  \quad \quad ; \quad \mbox{all other
components of $t_{ij}^k$ vanish} \label{Tvecti}
\end{eqnarray}
In the standard nomenclature, homogeneous cosmologies based on the
algebra (\ref{MaurerCartan}) are named  \textit{spaces of Bianchi
type 2A}. We shall prove that all possible oxidations of the
$\mathrm{A_2}$ sigma model solutions found in
\cite{piervoiastatia} are in the same $\mathrm{U}$--orbit together
with a purely gravitational representative corresponding to a
$D=4$ {Bianchi type 2A} space times a $\mathrm{T^6}$ torus or a
reduction thereof. More specifically we have Bianchi type 2A
spaces that are exact solutions for either the dilaton gravity
lagrangian in $d=4$
\begin{equation}
  A_{\mbox{dilaton gravity}}= \int \,  \sqrt{-\mbox{det} \, g} \, \left \{2 \, R[g] \,
  - \ft 12 \, \partial^\mu \phi \, \partial_\mu \phi   \right \}
\label{dilagravact}
\end{equation}
or for the  $0$--brane $d=4$ lagrangian
\begin{equation}
  A_{\mbox{0-brane}}= \int \,  \sqrt{-\mbox{det} \, g} \, \left \{2 \, R[g] \,
  - \ft 12 \, \partial^\mu \phi \, \partial_\mu \phi + \ft 1 4 \,
  \exp[-a \, \phi] \, F_{\mu\nu} \, F^{\mu\nu} \right \}
\label{1-branad4}
\end{equation}
This follows from what we are going to show shortly, namely
that, up to conjugation, there is just one \textit{regular embedding}
\begin{equation}
  {\bf i}:\mathrm{A_2} \, \, \hookrightarrow \, \mathrm{E_{8}}
\label{a2intoE8}
\end{equation}
This result has two relevant implications.  The first, which we
extensively discuss in the present paper, is that, through this
argument, we retrieve new exact solutions of Bianchi type 2A. The
second implication is that, via $\mathrm{U}$--transformations,
such new solutions, whose potential  interest in cosmology is
evident, are dual to solutions made of $D$-branes and admitting a
microscopic description in terms of time--dependent boundary
states. Indeed in \cite{piervoiastatia} we already oxided the
$\mathrm{A_2}$ sigma model solution to a $D=10$ configuration
containing a diagonal metric plus a system composed of a $D3$
space--brane and a $D1$ space--string. Due to the uniqueness of
the orbit, Bianchi cosmology 2A and this system are dual.
\par
Our method to study $\mathrm{U}$--orbits is based on the following
preliminary steps. Thanks to the techniques developed in
\cite{piervoiastatia} we know that each solution is obtained in
the following way. Let $\mathbf{G}_r \subset E_{8(8)}$ be a
regularly embedded maximally non compact subalgebra of the
$\mathrm{U}$--duality algebra of rank $r$. Let $\mathbf{H}_r
\subset \mathbf{G}_r$ be its maximal compact subalgebra which is
also necessarily contained in the maximal compact subalgebra of
$E_{8(8)}$ namely $\mathbf{H}_r \, \subset \, \mathrm{SO(16)}$.
Let furthermore $\mathcal{H}_r \subset \mathbf{G}_r$ be the Cartan
subalgebra of the chosen $\mathbf{G}_r$ which is necessarily a
subalgebra $\mathcal{H}_r\subset \mathcal{H}_8$ of the $E_{8(8)}$
Cartan subalgebra. The $\mathbf{G}_r$-solution is obtained from a
generating solution that lies only in the Cartan subalgebra
$\mathcal{H}_r$ by means of a compensating $\mathbf{H}_r $
transformation.  The $\mathrm{U}$--orbit of the
$\mathbf{G}_r$-solution is given by the orbit of possible regular
embeddings:
\begin{equation}
{\bf i} \, : \,  \mathbf{G}_r \hookrightarrow \mathrm{E_{8}}
\label{Grorbit}
\end{equation}
To this effect it is necessary and sufficient to restrict one's
attention to the discrete Weyl group $\mathcal{W} \subset
\mathrm{U}$. Indeed $\mathcal{W}$ maps the Cartan subalgebra into
itself and permutes the set of roots. $\mathrm{U}$--duality orbits
of $\mathbf{G}_r$ solutions are just the Weyl orbits of the
$\mathbf{G}_r$ root system.
\par
Therefore in this paper we perform a systematic study of the Weyl
orbits of $E_{8}$ regular subalgebras. We develop an algorithm
for such a study and we explore the nested chain of
$\mathrm{A_r}$ subalgebras of increasing rank. This chain is
particularly important since its canonical realization is within
the $\mathrm{A_{6}} \subset \mathrm{E_{8}}$ subalgebra that
describes the metric moduli of the $T^7$ torus in the reduction
from $D=10$ to $D=3$ dimensions. Hence any $\mathrm{A_r}$
solution, if the Weyl orbit is unique, has a purely gravitational
realization which is dual to other realizations eventually made
out of branes. Indeed, from our analysis it appears that there is
just one Weyl orbit for $\mathrm{A_2}$, $\mathrm{A_3}$
up to $\mathrm{A_{6}}$ subalgebras. Beyond we have two orbits
corresponding to the type IIA and type IIB interpretations of the
theory.
\par
In particular the uniqueness of the $\mathrm{A_2}$ orbit is
the proof of what we claimed above. \textit{Bianchi type 2A
cosmologies are dual to suitable space--brane systems.}
\par
Our paper is organized as follows:
\par
In section \ref{wile} we study the algebraic setup to classify Weyl
orbits of regular subalgebras, we outline the physical implications
of this classification and we derive our general algebraic results.
\par
In section \ref{canone} we study the canonical purely metric
representative of the $\mathrm{A_2}$ Weyl orbit and we show
that it is related to exact solutions of homogeneous Bianchi
cosmologies based on the Heisenberg algebra.
\par
In section \ref{bianchi} we perform a detailed study of the new exact
solutions of Bianchi type 2A that we have obtained through our
construction.
\par
Section \ref{conclu} contains our conclusions and perspectives.
\section{Weyl orbits of $E_{8(8)}$ subalgebras and Oxidation}
\label{wile}
As we anticipated in the introduction, the problem of classifying
different oxidations of the same $\sigma$--model solutions is reduced
first to the classifications of embeddings (\ref{Grorbit}) and then
to the classification of Weyl orbits of the $\mathbf{G}_{r}$ root
system within the $E_{8(8)}$ root system.\footnote{The action of the Weyl group
as a mean to permute field strengths in various $p$--brane solutions was already considered years ago in \cite{lupope}.}
\par
Let us now review the physical arguments leading to this conclusion.
\par
Using the solvable Lie algebra representation of the
$\mathrm{E_{8(8)}/SO(16)}$ coset manifold \cite{solv}, every
one--parameter (time) dependent solution of the $\sigma$--model is
described by the map:
\begin{equation}
\mathbb{R}\ni t \, \mapsto \, \left\{ h^i(t) \, , \,
\phi^\alpha(t)\right\}\, \equiv \, \Phi(t) \, \in \, Solv\left( E_8\right)
\label{mappaR}
\end{equation}
where $h^i(t)$ are the fields associated with the CSA generators,
$\phi^\alpha(t)$ the fields associated with the positive roots $\alpha
>0$ and altogether $\Phi(t)$ is a map of the time--line into the
Borel subalgebra $Solv\left( E_8\right) \subset E_{8(8)}$.
\par
Oxidation is a uniquely identified procedure that maps $\Phi(t)$ into
a solution of either type IIA or type IIB supergravity:
\begin{equation}
  Ox \, : \, \Phi(t) \, \mapsto \, F(t) = \mbox{supergravity solution}
\label{Oxidation}
\end{equation}
According to the results of \cite{piervoiastatia} a systematic
search of the possible time dependent backgrounds is performed in
the following way. First single out a maximally non compact
regularly embedded subalgebra ${\bf i} \left[
\mathbf{G}_{r}\right]  \subset E_{8(8)}$ and let ${\bf i} \left
[\mathbf{H}_r \right]\subset {\bf i} \left [\mathbf{G}_{r}\right
]$ be its compact subalgebra. Here ${\bf i}$ denotes the specific
embedding, while $\mathbf{G}_r$ denotes the abstract algebra. The
pair $\left \{\mathrm{G}_r \, , \, \mathrm{H}_r \right \}$ defines
a new $\sigma$--model with target space
$\mathrm{G}_{r}/\mathrm{H}_r$ which also admits a solvable Lie
algebra description. Hence  every one--parameter (time) dependent
solution of this new (smaller) $\sigma$--model is described by a
map similar to that in (\ref{mappaR}):
\begin{equation}
\mathbb{R}\ni t \, \mapsto \, \left\{ h_{G_r}^i(t) \, , \,
\phi_{G_r}^\alpha(t)\right\}\, \equiv \, \Phi_{G_r}(t) \, \in \, Solv\left( \mathbf{G}_r\right)
\label{mappaRG}
\end{equation}
Each different embedding (\ref{Grorbit})  defines a different explicit
solution of the $\mathrm{E_{8(8)}/SO(16)}$ sigma model:
\begin{equation}
  {\bf i} \left[\Phi_{G_r}(t)\right ] \, = \, \Phi_{{\bf i}}(t)
\label{iotaPhi}
\end{equation}
which through oxidation (\ref{Oxidation}) leads to a different
supergravity background.
\par
Let us now recall the algorithm developed in \cite{piervoiastatia}
in order to obtain general integrals of the
$\mathrm{G}_r/\mathrm{H}_r$ sigma--model differential equations.
There we showed that these equations have the structure of
geodesic equations for the scalar manifold
$\mathrm{G}_r/\mathrm{H}_r$ and that one can first obtain a
generating solution, corresponding to a \textit{normal form}
orientation of the geodesic tangent vector in the origin ($t=0$).
Such a normal form orientation is provided by a tangent vector
pointing only in the direction of the Cartan generators and can be
reached by means of $\mathrm{H}_r$ transformations. Indeed the
isotropy group has a linear action on the tangent space to the
target manifold. The corresponding generating solution has the
form:
\begin{equation}
  \Phi^{(0)}_{G_r}(t) =\left\{ \overline{h}_{G_r}^i(t) \, , \,
0\right\} \quad ; \quad  \overline{h}_{G_r}^i(t) = \exp [\omega^i \,
t + c^i]
\label{normalaforma}
\end{equation}
A full solution where also the \textit{root fields} are excited is
obtained from the generating solution by means of compensating
$\mathrm{H}_r$--transformations, namely $\mathrm{H}_r$
transformation with parameters $\theta_{G_r}(t)$ such that a
solvable parametrization of the coset $\mathrm{G}_r/\mathrm{H}_r$
is mapped into another solvable parametrization. Such a condition
is guaranteed by a system of differential equations which is
equivalent to the original system to be solved but has the
advantage of being already in triangular form and hence reduced to
quadratures. Schematically we have:
\begin{equation}
  \Phi_{G_r}(t) = H\left(\theta_{G_r}^\alpha(t)
  \right) \cdot \Phi^{(0)}_{G_r}(t)  \quad \Rightarrow \quad
  {\bf i} \left[\Phi_{G_r}(t)\right] = {\bf i} \left[
  H\left(\theta_{G_r}(t) \right ) \right] \cdot {\bf i}\left[\Phi^{(0)}_{G_r}(t)\right]
\label{fromgener}
\end{equation}
where $H\left(\theta_{G_r}(t) \right)$ is the compensating
transformation. In (\ref{fromgener}) the first is the relation in
the abstract algebra $\mathbf{G}_{r}$, while the second is its
realization in the specific subalgebra  ${\bf i} \left [
\mathbf{G}_{r} \right] \subset \mathrm{E_{8}}$. Using these
notations we can address the structure of a generic $E_{8(8)}$
transformation that maps one supergravity solution $F_1^{G_r}(t)$
associated with one embedding ${\bf i}_1$ of the Lie algebra
$\mathbf{G}_r$ to another supergravity solution $F_2^{G_r}(t)$
associated with a second embedding ${\bf i}_2$. It suffices to
note that for any two regular embeddings of the same algebra we
have:
\begin{equation}
  {\bf i}_1 \, \left [ \mathbf{G}_{r} \right ] \, = \, \mathbf{w}_{12}
  \, {\bf i}_2 \, \left [ \mathbf{G}_{r} \right ] \quad \mbox{where} \quad
  \mathbf{w}_{12} \, \in \, \mathcal{W} \left ( \mathrm{E_{8}}\right )
\label{weyltransfa}
\end{equation}
having denoted by $\mathcal{W} \left ( \mathrm{E_{8(8)}}\right )$ the
Weyl group of the $E_{8(8)}$ algebra. In eq. (\ref{weyltransfa}) the
action of the Weyl group element $\mathbf{w}_{12}$ is the natural one
on the algebra.
Then we get
\begin{equation}
  \Phi_{{\bf i}_1[\mathbf{G}_r]}(t) = \, \mathbf{w}_{12}
  \, \Phi_{{\bf i}_2[\mathbf{G}_r]}(t)
\label{urcave}
\end{equation}
Fig.(\ref{disegcom}) summarizes our understanding of this Weyl
mapping between different supergravity backgrounds.
\iffigs
\begin{figure}
\begin{center}
\epsfxsize =11cm {\epsffile{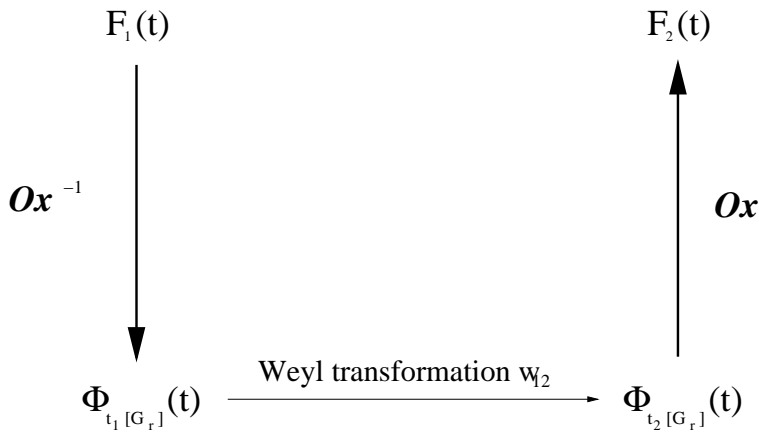} \vskip -3.5cm
\caption{\label{disegcom} The Weyl group $\mathcal{W}(E_{8})$ maps
different embeddings of the same subalgebra $G_{r}$ into
$E_{8(8)}$. There is just one abstract solution of the
$\mathrm{G}_{r}/\mathrm{H}_r$ sigma model that leads to many
different physical oxidations related one to the other by Weyl
transformations.} } \hskip 2cm \unitlength=1.1mm
\end{center}
\end{figure}
\fi
\par
From this argument it follows that it is of primary relevance to
study orbits of $E_{8(8)}$ subalgebras under the Weyl group. Since
each embedding corresponds to a different oxidation, namely to a different
$D=10$ interpretation of the same abstract solution, it follows that all oxidations in the same orbit
are equivalent under duality transformations.
\subsection{Preliminaries on the Weyl group}
Every regular embedding of a smaller semisimple subalgebra
$\mathbf{G}_r$ into a bigger one ($\mathrm{E_{8}}$ in our case) is
uniquely specified by the embedding of the small CSA
$\mathcal{H}_r \subset \mathbf{G}_r$ into the  big one
$\mathcal{H}_r \hookrightarrow \mathcal{H}_8$. The embedding of
the roots follows uniquely, once the embedding of the CSA is
given. Vice versa, the choice of the simple roots of the
subalgebra $\mathbf{G}_{r}$ inside the root system of the big
algebra $E_{8(8)}$ fixes the embedding of the CSA, the relevant
map being
\begin{equation}
  \alpha\rightarrow h_{\alpha}
\label{pracitat}
\end{equation}
So, any embedding is specified by a set of Cartan generators associated with the
simple roots of the algebra $\mathbf{G}_r$ to be embedded:
\begin{equation}
  {h_{\alpha_s}} = h_i\cdot\alpha_s^i, \quad s=1,
\dots, r
\label{embeddingCSA}
\end{equation}
where ${h_i}, \, i=1,\dots 8$ denote  any suitable
basis of the Cartan subalgebra of $\mathrm{E_{8(8)}}$.
To convert one embedding into another, we just have the Weyl group, which, by definition, is generated by
the reflections $\sigma_\gamma$ with respect to the plane orthogonal to any of the
roots $\gamma \in \Delta$:
\begin{equation}
\sigma_{\gamma}\, {\alpha} = \alpha - \langle \alpha \, , \,
\gamma \rangle  \, \gamma \quad ;
 \quad \langle \alpha \, , \, \gamma \rangle  \, \equiv \, 2 \frac{\alpha \cdot \gamma}{ \alpha \cdot \alpha}
 \label{rfleczia}
\end{equation}
\subsection{Strategy to classify orbits}
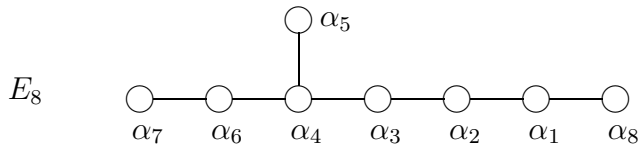
\begin{figure}
\centering
\begin{picture}(100,100)
        \put (-70,35){$E_8$} \put (-20,35){\circle {10}} \put
(-23,20){$\alpha_7$} \put (-15,35){\line (1,0){20}} \put
(10,35){\circle {10}} \put (7,20){$\alpha_6$} \put (15,35){\line
(1,0){20}} \put (40,35){\circle {10}} \put (37,20){$\alpha_4$}
\put (40,65){\circle {10}} \put (48,62.8){$\alpha_5$} \put
(40,40){\line (0,1){20}} \put (45,35){\line (1,0){20}} \put
(70,35){\circle {10}} \put (67,20){$\alpha_{3}$} \put
(75,35){\line (1,0){20}} \put (100,35){\circle {10}} \put
(97,20){$\alpha_{2}$} \put (105,35){\line (1,0){20}} \put
(130,35){\circle {10}} \put (127,20){$\alpha_1$} \put
(135,35){\line (1,0){20}} \put (160,35){\circle {10}} \put
(157,20){$\alpha_8$}
\end{picture}
\vskip 1cm \caption{The Dynkin diagrams of $E_{8(8)}$ and the labeling of simple roots}
\label{except2}
\end{figure}

In this way we have concluded that the main algebraic question to be
answered is how to connect by means of Weyl transformations  different
choices of the simple roots (or of the CSA subalgebra) of a given
abstract $\mathbf{G}_r$ inside the root system of $E_{8}$. To
illustrate our strategy for the solution of such a problem  we begin by presenting the
Dynkin diagram of $\mathrm{E_{8}}$, which is done in
fig.\ref{except2}.
This fixes our conventions and notations for simple roots, which
are those of \cite{piervoiastatia}. In that paper we also
presented a table listing all the $120$ positive roots, arranged
by height and ordered according to our conventions. We will
frequently refer to such table for the absolute identification of
the roots $\alpha[i]$ by means of a number $i$ ranging from $1$ to
$120$. The correspondence between such a number and the Dynkin
labels, according to the naming of fig.\ref{except2}, is given in
the aforementioned table of \cite{piervoiastatia}.
\par
 The strategy we adopt in classifying regular embeddings of $\mathbf{G}_r$ in
terms of Weyl orbits can be summarized as follows. We consider a
set of linearly  independent  roots ${\gamma_1,\dots, \gamma_r}$
of $\mathbf{G}_r$, and embed them sequentially within the root
system of $\mathrm{{E}_{8(8)}}$ by requiring that the
transformations which we use to fix $\gamma_{i+1}$ should belong
to the stability subgroup $I^{\gamma_i}_S \subset \mathcal{W}({\rm
E}_{8})$ which, inside the Weyl group,  leaves the previously
fixed roots $\gamma_1,\dots, \gamma_i$ invariant. Let us list
mathematical properties
 of the Weyl group which we find convenient to recall at this junction.
They are proved in many standard textbooks (see for instance
\cite{humphrey2}):
\par
\begin{itemize}
\item all roots of the $\mathrm{E_8}$ Lie algebra lie just in one
orbit under the  action of the Weyl group
$\mathcal{W}(\mathrm{E_8})$. And, in particular, with the help of
the Weyl group $\mathcal{W}(\mathrm{E_8})$ we can map any root
into the highest root $\alpha[120]$. \item the stability subgroup
of any root of $\mathrm{E_8}$ under action of the Weyl group
$\mathcal{W}(\mathrm{E_8})$ is isomorphic to Weyl group
$\mathcal{W}(\mathrm{E_7})$. Indeed the roots orthogonal to the
highest root $\alpha[120]$ are just those whose component $n_8$
vanishes and these are the roots of $\mathrm{E_7}$, \item weights
of fundamental representation of $\mathrm{E}_r$ constitute just
one orbit under the action of $\mathcal{W}(\mathrm{E}_r)$ for
$r=4,5,6,7$.
\end{itemize}
Keeping the above facts ready for use we can develop the embedding
procedure we sketched above. According to it we encounter a {\it
branching} in the Weyl--orbits for the $\mathbf{G}_r$ Lie algebra
embeddings when we there is more than one possible  orbit of
$I^{\gamma_i}_S$ in which  to choose the root $\gamma_{i+1}$.
\par
\subsubsection{Orbits of $\mathrm{A}_r$ algebras}
We illustrate our method by considering the embeddings of the
$\mathbf{G}_r=\mathrm{A_r}$ algebras, $r=1,\dots, 8$. Here we
choose as representative roots $\gamma_i$ the set of $r$ linearly
independent roots arranged in order of decreasing height starting
from $\gamma_1$ equal to the highest root of $\mathrm{A_r}$ and
all the other chosen in such a way that we always have:
\begin{equation}
\forall i \ne j \quad : \quad   \gamma_i\cdot \gamma_j=1
\label{prillone1}
\end{equation}
In intrinsic Dynkin labels of the $\mathrm{A_r}$ Lie algebra, the
roots satisfying the constraint (\ref{prillone1}) are the following
ones:
\begin{eqnarray}
\begin{array}{cccccccc}
  \gamma_1 & = & (1,1,1,\dots,1) & ; & \gamma_2 & = & (1,1,\dots,1,0) & \dots \\
  \gamma_{r-1} & = & (1,1,0,\dots,0,0) & ; & \gamma_r & = & (1,0,\dots,0,0) &
  .
\end{array}
\label{reprebeta}
\end{eqnarray}
\begin{itemize}
\item{{\it Choosing $\gamma_1$:} since ${\rm Adj}({\rm E}_{8(8)})$
contains a single $\mathcal{W}({\rm E}_{8})$--orbit, all choices
of $\gamma_1$ are connected by Weyl transformations. Moreover the
stability group of any ${\rm E}_{8(8)}$ root is
$I^{\gamma_1}_S=\mathcal{W}({\rm E}_{7})$. With respect to ${\rm
E}_{7(7)}\times {\rm O}(1,1)$  the adjoint of  ${\rm E}_{8(8)}$
branches as follows:
\begin{eqnarray}
{\bf 248}&\rightarrow & {\bf 133}_0+{\bf 56}_{+1}+{\bf 56}_{-1}+{\bf 1}_{+2}+{\bf 1}_{-2}+{\bf 1}_{0}
\end{eqnarray}
 The root $\gamma_1$ coincides with ${\bf 1}_{+2}$ with respect to its own stability group.}
\item{{\it Choosing $\gamma_2$:} } the only possible orbit of
$I^{\gamma_1}_S=  \mathcal{W}({\rm E}_{7})$, where  a root,
defined by a scalar product $\gamma_1\cdot\gamma_i = 1$, can lie
is the representation ${\bf 56}_{+1}$. Since all the remaining
roots $\{\gamma_i\}, \quad i=2,\dots,r$, of our set have such a
property, all the others also belong to ${\bf 56}_{+1}$.  All
members of this orbit are connected by Weyl transformations of
$\mathcal{W}({\rm E}_{7})$. The stability group of any root
(weight) of the ${\bf 56}_{+1}$ is $I^{\gamma_2}_S=
\mathcal{W}({\rm E}_{6})$ (this is the stability group for a pair
of roots of ${\rm E}_{8(8)}$, which have scalar product equal to
one). With respect to ${\rm E}_{6(6)}\times {\rm O}(1,1)$ the
adjoint of ${\rm E}_{7(7)}$ and the representation ${\bf 56}_{+1}$
decompose, respectively:
\begin{eqnarray}
\nonumber  {\bf 133}&\rightarrow & {\bf 78}_0 + {\bf 1}_{0}+{\bf 27}_{-2}+{\bf \overline{27}}_{+2}, \\
{\bf 56}&\rightarrow & {\bf 1}_{+3}+{\bf 27}_{+1}+{\bf
1}_{-3}+{\bf \overline{27}}_{-1}
\end{eqnarray}
 The root $\gamma_2$ is the singlet ${\bf 1}_{+3}$.
 \item{{\it Choosing $\gamma_3$:}} The roots $\gamma_i, \quad i=3,\dots, r$, are identified as the roots
 belonging to ${\bf 56}_{+1}$ which satisfy the additional condition $\gamma_2\cdot\gamma_i = 1$. The only orbit of
 $I^{\gamma_2}_S= \mathcal{W}({\rm E}_{6})$, where such roots can be found, is given by the weights of the
 fundamental representation ${\bf 27}_{+1}$.  The stability group of this
 representation is $I^{\gamma_3}_S= \mathcal{W}({\rm E}_{5})$.  With respect to ${\rm E}_{5(5)}\times {\rm O}(1,1)$
 the adjoint of ${\rm E}_{6(6)}$ and the representation ${\bf  27}_{+1}$ decompose, respectively:
\begin{eqnarray}
\nonumber  {\bf 78}&\rightarrow & {\bf 45}_0 + {\bf 1}_{0}+{\bf 16}_{-3}+{\bf \overline{16}}_{+3}, \\
{\bf 27}&\rightarrow & {\bf 1}_{+4}+{\bf 16}_{+1} + {\bf 10}_{-2}
\end{eqnarray}
 The root $\gamma_3$ is necessarily fixed to be the singlet ${\bf 1}_{+4}$, with respect to its stability subgroup.
  \item{{\it Choosing $\gamma_4$:}} the remaining roots $\gamma_i, \quad i=4,\dots,r$, are inside ${\bf 27}_{+1}$ and
have the scalar product $\gamma_3\cdot\gamma_i = 1$ with the
singlet ${\bf 1}_{+4}=\gamma_3$. The only orbit of $I^{\gamma_3}_S
=\mathcal{W}({\rm E}_{5})$, where such a root can be, is ${\bf
16}_{+1}$. The stability group of this representation is
$I^{\gamma_4}_S= \mathcal{W}({\rm SL}(5))$.  With respect to ${\rm
SL}(5)\times {\rm O}(1,1)$ the adjoint of ${\rm E}_{5(5)}$ and the
representation ${\bf 16}_{+1}$ decompose, respectively, as
follows:
\begin{eqnarray}
\nonumber  {\bf 45}&\rightarrow & {\bf 24}_0 + {\bf 1}_{0}+{\bf 10}_{-4}+{\bf \overline{10}}_{+4}, \\
{\bf 16}&\rightarrow & {\bf 1}_{+5}+{\bf 10}_{+1} + {\bf
\overline{5}}_{-3}
\end{eqnarray}
 We fix $\gamma_4$ to be a singlet ${\bf 1}_{+5}$.
  \item{{\it Choosing $\gamma_5$:}} the roots $\gamma_i, \quad i=5,\dots,r$, belong to ${\bf 16}_{+1}$ and
have a scalar product $\gamma_4\cdot\gamma_i = 1$ with the singlet
${\bf 1}_{+5}$. The only orbit of $I^{\gamma_4}_S =
\mathcal{W}({\rm SL}(5))$, that we can choose at this step, is
${\bf 10}_{+1}$. The stability group of this representation is
$I^{\gamma_5}_S=\mathcal{W}({\rm SL}(3)\times{\rm SL}(2))$.  With
respect to ${\rm SL}(3)\times{\rm SL}(2)\times {\rm O}(1,1)$ the
adjoint of ${\rm SL}(5)$ and the representation ${\bf 10}_{+1}$
decompose, respectively, as follows:
\begin{eqnarray} \nonumber  {\bf 24}&\rightarrow & {\bf (8,1)}_0 + {\bf (1,3)}_{0}+
{\bf (1,1)}_{0}+{\bf (3,2)}_{-5}+{\bf (\overline{3},2)}_{+5}, \\
{\bf 10}&\rightarrow & {\bf (1,1)}_{+6}+{\bf (3,2)}_{+1} + {\bf
(\overline{3},1)}_{-4}\end{eqnarray}
 We fix $\gamma_5$ to be a singlet ${\bf (1,1)}_{+6}$.
  \item{{\it Choosing $\gamma_6$:}} the roots $\gamma_i, \quad i=6,\dots,r$, belong to
  ${\bf (3,2)}_{+1}$.
The stability group of this representation is $I^{\gamma_6}_S=
\mathcal{W}({\rm SL}(2)\times{\rm O}(1,1))$. With respect to ${\rm
SL}(2)\times{\rm O}(1,1)\times {\rm O}(1,1)$ the adjoint of ${\rm
SL}(3)\times{\rm SL}(2)$ and the representation ${\bf (3,2)}_{+1}$
decompose, respectively:
\begin{eqnarray}
{\bf (8,1)} &\rightarrow & {\bf 3}_{(0,0)}+ {\bf 1}_{(0,0)}+{\bf
2}_{(-3,0)}+
{\bf 2}_{(+3,0)}\nonumber\\
{\bf (1,3)} &\rightarrow & {\bf 1}_{(0,+14)}+ {\bf
1}_{(0,-14)}+{\bf 1}_{(0,0)}\nonumber\\
{\bf (3,2)}&\rightarrow & {\bf 1}_{(+2,+7)}+{\bf 1}_{(+2,-7)}+
{\bf 2}_{(-1,-7)}+ {\bf 2}_{(-1,+7)}\end{eqnarray}
 $\gamma_6$ is one of the two singlets, e.g. ${\bf 1}_{(+2,+7)}$.
\item{{\it Choosing $\gamma_7$:}} the root $\gamma_7$ belongs to
${\bf (3,2)}_{+1}$ and has scalar product one with the singlet
${\bf 1}_{(+2,+7)}$. We have {\textit{two possible choices}} for
$\gamma_7$: either the other singlet ${\bf 1}_{(+2,-7)}$ or the
doublet ${\bf 2}_{(-1,+7)}$. These choices define distinct Weyl
orbits for $A_7$, since they can not be mapped into each other by
the action of the stability group $I^{\gamma_6}_S
=\mathcal{W}({\rm SL}(2)\times{\rm O}(1,1))$. This is an instance
of the branching mentioned above which implies that not all the
$\mathrm{A_7}$ models can be mapped one into the other by means of
$\mathcal{W}({\rm E}_{8})$, indeed they fall into two distinct
orbits, as contrary to the smaller rank cases $\mathrm{A}_{r<7}$
which fall into single orbits. The two orbits of
$\mathrm{A_7}$--embeddings have different stability groups: if we
choose $\gamma_7$ as the singlet ${\bf 1}_{(+2,-7)}$, the
stability group of the corresponding Weyl--orbit is
$I^{\gamma_7}_S=\mathcal{W}({\rm SL}(2))$, while if we choose
$\gamma_{7\prime}$ inside the doublet ${\bf 2}_{(-1,+7)}$ then
$I^{\gamma_7^\prime}_S=\mathcal{W}({\rm O}(1,1))=\varnothing$.
\item{{\it Choosing $\gamma_8$:}} If  $\gamma_7$ is chosen to be
the singlet ${\bf 1}_{(+2,-7)}$ then it is straightforward to show
that none of the representations of
$I^{\gamma_7}_S=\mathcal{W}({\rm SL}(2))$ can contain $\gamma_8$.
Therefore $A_7$ representatives of this orbit can not be further
extended to $A_8$. On the other hand, if $\gamma_{7\prime}$ had
been chosen inside the doublet ${\bf 2}_{(-1,+7)}$ then the root
$\gamma_8$ can be taken to be the other element of the doublet
which thus defines the right orbit of $I^{\gamma_{7\prime}}_S$. At
this stage all the original Weyl group is completely broken,
namely $I^{\gamma_8}_S=\varnothing$.
\end{itemize}
Summarizing, we have seen that all regular embeddings of each
$A_r$ subalgebra inside ${\rm E}_{8(8)}$ fall into a single Weyl
orbit, except $A_7$ which fall into two distinct ones.
\par
\subsubsection{Interpretation of the $\mathrm{A}_r$ orbits from the
dimensional oxidation viewpoint} There is an interesting
interpretation  of the $\mathrm{A}_r$--orbits and of their
stability groups in terms of dimensional oxidation from $D=3$ to
$D=3+r$. We may identify a representative of the $\mathrm{A}_r$
algebra inside each orbit as generating the group which acts
transitively on the metric moduli of an internal $\mathrm{T^r}$
torus $G_{ij}$ ($i,j=1,\dots , r$) plus the dualized Kaluza--Klein
vectors in three dimensions $\gamma_\mu^{3+i}$, corresponding
respectively to the roots $\gamma_i$. In this case all the scalar
fields described by these models are singlets with respect to the
duality group $\mathrm{U_D}$ of $D=3+r$ maximal supergravity.
Consistently we find that the stability group of the
$\mathrm{A_r}$ embedding coincides with the automorphism group of
$U_{3+r}$. For $r=7$ the two Weyl orbits define the ${\rm SL}(8)$
group of the $\mathrm{T^7}$ metric moduli and Kaluza--Klein
vectors in Type $IIA$ or Type $IIB$ descriptions respectively. The
corresponding stability groups
$I^{\gamma_7^\prime)}_S=\mathcal{W}({\rm O}(1,1))$ and
$I^{\gamma_7}_S=\mathcal{W}({\rm SL}(2))$ are indeed related to
the $D=10$ duality groups of these two theories. We have seen that
only in the Type $IIA$ the embeddings can be further extended to a
unique $\mathrm{A_8}$ algebra, which is related to oxidation to
$M$--theory.
\par
The uniqueness of the Weyl orbit for the Lie algebras
$\mathrm{A}_r$ (for $r<7$) has the relevant implications we
already announced in the introduction. In the same orbit there are
several embeddings which involve various different set of fields,
$B$--fields, $RR$--fields and so on, but there is always one
canonical representative which involves only metrics $G_{ij}$ and
Kaluza--Klein vectors, namely also metrics one dimension above.
Hence purely metric configurations are dual to configurations
which involve $RR$--fields and can be described in terms of
branes.
\par
Solutions defined by a semisimple Lie algebra ${\bf G}_r$ different
from $\mathrm{A_r}$  will also have an interpretation in terms of
oxidation to higher dimensions, however their scalar fields will not be related
just to the metric moduli.
\par
We postpone the construction and classification of other algebra
orbits to a future publication.
\section{The canonical, pure metric representative of the $\mathrm{A_2}$ orbit}
\label{canone} In paper \cite{piervoiastatia} we constructed the
general integral for an $\mathrm{A_2}$ model, namely for the
abstract sigma model over the target manifold
$\mathrm{SL(3)/\mathrm{SO(3)}}$. In the same paper we studied
oxidation of such abstract solutions to $D=10$. That involved
choosing an embedding $\mathrm{A_2} \hookrightarrow
\mathrm{E_{8(8)}}$ and we chose the following one\footnote{In the
conventions of \cite{piervoiastatia} that we follow here the
simple root $\alpha[7]$ is the spinorial root of $D_6$, namely
$\alpha[7]= -\ft 12 \sum_{i=1}^8\epsilon_i$, while $\epsilon_i$
are the unimodular
 orthonormal vectors in $\mathbb{R}^8$}:
\begin{equation}
  \begin{array}{cccclcc}
          {\bf i}[\beta_1]& = &  \alpha [69] & = &\epsilon _1 + \epsilon _2 &  \leftrightarrow & B_{34}\\
          {\bf i}[\beta_2] & = & \alpha [15] & = & \alpha[7] +\epsilon _6 + \epsilon _7  &
          \leftrightarrow & C_{89}  \\
          {\bf i}[\beta_3]& = & \alpha [80] & = & \alpha [7] +\epsilon _1 + \epsilon _1 +
          \epsilon _6 + \epsilon _7&  \leftrightarrow & C_{3489} \sim C_{\mu 567} \
  \end{array}
\label{identificazie}
\end{equation}
where  $\beta_{1,2}$ are the simple roots of the
$\mathrm{A_2}$ Lie algebra and $\beta_3 =\beta_1 + \beta_2$
is the highest root. The corresponding oxided solutions of type
IIB supergravity describe a system with a diagonal metric which
contains both a space $D3$--brane in the $3489$ directions and a
space $D1$--brane in the directions $89$.
\par
From the results of the previous section we know that this
embedding is in the same Weyl orbit together with a canonical
representative ${\bf i}_{can}[\mathrm{A_2}]$, which is purely
metric and which, following the procedure outlined above, can be
explicitly retrieved. The highest root $\beta_1 + \beta_2$ is
identified with the highest root of $\mathrm{E_{8(8)}}$, namely
with $\alpha[120]$, while $\beta_2$ is identified with the root
next to highest $\alpha[119]$. Altogether we have:
\begin{equation}
  \begin{array}{cccclcc}
        {\bf i}_{ {can}} [\beta_1 ]& = &   \alpha[8] & = &\epsilon_1 -
 \epsilon_2 &  \leftrightarrow & \gamma_{4}^3\\
            {\bf i}_{ {can}}[ \beta_2]  & = & \alpha[119] & = & \epsilon_2 - \epsilon_8  &
          \leftrightarrow &\gamma_{\mu}^4  \\
          {\bf i}_{ {can}}[ \beta_3]& = &  \alpha[120] & = & \epsilon_1 - \epsilon_8&
          \leftrightarrow & \gamma_{\mu}^3 \
  \end{array}
\end{equation}
It is of the utmost interest to explore the properties of this
canonical representative. It will turn out that it provides examples
of metrics in $D=4$ which fall in the general class of homogeneous
cosmologies classified by Bianchi more than 80 years ago. More specifically
it provides exact solutions for  Bianchi type 2A metrics, associated
with the Heisenberg algebra, as defined by the Maurer Cartan equations
in eq.(\ref{MaurerCartan}).
\par
Let us see how this happens in the explicit process of oxidation.
\par
Given the explicit form of the three roots we immediately see that
the Cartan subalgebra of this ${\bf i}_{can}[\mathrm{A_2}]$
model is spanned by
 all 8--vectors of the following form:
\begin{equation}
 {\bf i}_{can}  [\vec{h} ] = \{x,y,0,0,0,0,0,-x-y\}
\end{equation}
To express $x$ and $y$ in terms of $\vec{h} = \left \{{h}_1, {h}_2
\right \}$, namely in terms of the Cartan scalar fields of the
abstract $\mathrm{A_2}$ model, we use ${\bf i}_{can} [
\vec{h}  ]\cdot\vec{\alpha}[8] = \vec{ h}\cdot\vec{\beta}_1$,
${\bf i}_{can} [ \vec{h}  ]\cdot\vec{\alpha}[119] = \vec{
h}\cdot\vec{\beta}_2$ and we find
 \begin{equation}
 x = \frac{1}{\sqrt{2}}h_1 + \frac{1}{\sqrt{6}}h_2, \quad y = -\frac{1}{\sqrt{2}}h_1
 + \frac{1}{\sqrt{6}}h_2
 \end{equation}
 The ten-dimensional dilaton in this embedding is zero, since
 \begin{equation}
 \phi = -{\bf i}_{can} [ \vec{h}  ]\cdot\vec{\alpha}[7] = 0
 \end{equation}
Then we proceed with the construction of the internal metric. By
definition, it is given by the product of the vielbein with the
transposed vielbein $g = E E^T$, where $E \mathcal{N}\mathcal{H}$, is the product of a diagonal matrix
$\mathcal{H} = \exp[\sigma_i]\delta_{ij}$ with the matrix
$\mathcal{N}$, which is exponential of the nilpotent generators.
The fields $\sigma_i$, $i = 1,\dots,7$, that correspond to the
radii of the internal metric, are obtained from
\begin{equation}
 {\bf i}_{can} [ \vec{h}  ] = \sum_{i=1}^7\sigma_i\epsilon_i + 2\phi_3\epsilon_8
\end{equation}
 So, we get
\begin{eqnarray}
 && \nonumber \sigma_1 = x, \quad \sigma_2 = y, \quad \sigma_i = 0, \quad i=3,\dots,7 \\
 && \phi_3 = -\ft 12(x+y)
\end{eqnarray}
From this identification we see that the metric we are going to
construct will be dynamical only in five dimensions. So, we can
think of this embedding as of an oxidation of the sigma-model
solutions to a pure metric configuration in five dimensions times
the metric of a straight $T^5$-torus \be ds^2_{T^5}=
\sum_{i=5}^9dx_i^2 \ee From now on we can consider the internal
metric to be 2-dimensional and represented by the $2 \times 2$
matrices\footnote{$\varphi_i$, $i=1,2,3$, are the scalar fields
associated with roots in the abstract $A_2$ model.}
\begin{eqnarray}
 \mathcal{H} = \left(\matrix{e^{x}& 0\cr 0 & e^y}\right), \quad \mathcal{N}  =\left(\matrix{1 & \varphi_1 \cr 0 & 1}\right)
\end{eqnarray}
so that the internal metric is finally given by:
\begin{eqnarray}
 g_{ij} = \left(\matrix{e^{2x}+\varphi_1^2 e^{2y}& \varphi_1 e^{2y}\cr \varphi_1e^{2y}& e^{2y}}\right)
\end{eqnarray}
The identification of the Kaluza-Klein vectors $\gamma_{\mu}^3$
and $\gamma_{\mu}^4$ in terms of the scalar fields $\varphi_i$
associated with the roots, involves a dualization procedure; the
result is
\begin{equation}
F^i_{\mu\nu} = \varepsilon_{\mu\nu 0}e^{4\phi_3}g^{ij}\dot{W}_j , \quad i,j = 3,4
\end{equation}
where
\begin{equation}W_3 = \ft 12\varphi_1\varphi_2 + \varphi_3, \quad W_4 = \varphi_2
\end{equation}
We can solve this dualization rule in terms of Kaluza-Klein vector potentials
as it follows:
\begin{equation}
\gamma^i_0 = 0, \quad \gamma^i_1 = -\ft12
x_2e^{4\phi_3}g^{ij}\dot{W}_j, \quad \gamma^i_2 = \ft12
x_1e^{4\phi_3}g^{ij}\dot{W}_j
\end{equation}
\subsection{Elaboration of the solution with one root switched on}
After fixing the identification of the embedding, we can oxide the
particular solutions we obtained in \cite{piervoiastatia}.
We start with the simplest one, the solution where only one root is switched
on. It reads as follows \cite{piervoiastatia}:
\begin{eqnarray}
h_1(t) & = &-\frac{t\kappa + 2
          \log ({\cosh(\frac{t\,\omega }{2})})}{4\,{\sqrt{2}}}, \hspace{0.5cm}
          h_2(t)  =  \frac{t\kappa  - 6\,\log (\cosh (\frac{t\,\omega }{2}))}{4\,{\sqrt{6}}},\nonumber\\
\varphi_1(t) &=& 0, \hspace{0.5cm} \varphi_2(t) = 0,
\hspace{0.5cm}
  \varphi_3(t) = \frac{{\sqrt{2}}}{1 + e^{t\,\omega }}
\label{finsol1}
\end{eqnarray}
and it oxides to the following 5-dimensional Ricci-flat metric
\begin{eqnarray}
 && \nonumber ds^2_5 = -e^{t\sqrt{\frac{\kappa^2}{3}+\omega} -
\frac{t\kappa}{6}}\cosh\frac{t\omega}{2} dt^2 + e^{\ft 12
t\sqrt{\frac{\kappa^2}{3}+\omega} -
\frac{t\kappa}{6}}\cosh\frac{t\omega}{2}\left(dx_1^2 +
dx_2^2\right) + \\ && +
\frac{e^{-\frac{t\kappa}{6}}}{\cosh\frac{t\omega}{2}}(dx_3 +
\frac{\omega}{4}(x_1dx_2 - x_2 dx_1))^2  + e^{\ft 23 \kappa
t}dx_4^2\label{ds5}
\end{eqnarray}
The metric that we obtain by putting $\kappa =0$ is, essentially,
4-dimensional and reads
\begin{equation}
ds^2_{\kappa=0} = -e^{t\omega}\cosh\frac{t\omega}{2} dt^2 +
e^{\frac{t\omega}{2}}\cosh\frac{t\omega}{2}\left(dx_1^2 +
dx_2^2\right)
 + \frac{1}{\cosh\frac{t\omega}{2}}\left ( dx_3 +
\frac{\omega}{4}(x_1dx_2 - x_2 dx_1)\right)^2 \label{ds4}
\end{equation}
As we extensively discuss in the next section, the metric in (\ref{ds4}) falls into the class
of Bianchi type 2A metrics and provides a remarkable example of exact
vacuum solution in that class, since it is exactly Ricci flat.
\subsubsection{$4D$--interpretation of the metric with $\kappa \ne 0$}
Switching on the parameter $\kappa$ leads to a non trivial evolution of the scale factor
also in the direction of $x^4$.
We can reinterpret this in $D=4$ by means of a standard Kaluza Klein
reduction of the metric (\ref{ds5}) on a $T^1$ torus, the compact  coordinate being precisely $x^4$.
From a $4$-dimensional point of view what has happened is that we have switched on a scalar field $\phi$,
corresponding to the metric component $g_{44} = \exp[\ft 13 \kappa t]$.
The dimensional reduction of the metric (\ref{ds5}) to four dimensions, according to the normalizations
of dilaton--gravity, as fixed by eq.(\ref{dilagravact}), yields:
\begin{equation}
\phi = \sqrt{\ft 3 2 }\, \log[\sqrt{g_{44}}] = \ft {1}{2\sqrt{6}} \kappa t
\label{linescalapre}
\end{equation}
for the scalar field and
\begin{eqnarray}
&& \nonumber ds^2_{E,4} -e^{t\sqrt{\frac{\kappa^2}{3}+\omega}}\cosh\frac{t\omega}{2} dt^2
+ e^{\ft 12 t\sqrt{\frac{\kappa^2}{3}+\omega}}
\cosh\frac{t\omega}{2}\left(dx_1^2 + dx_2^2\right) + \\ && +
\frac{1}{\cosh\frac{t\omega}{2}}(dx_3 + \frac{\omega}{4}(x_1dx_2 -
x_2 dx_1))^2 \label{dilametrapre}
\end{eqnarray}
for the 4-dimensional metric in the Einstein frame.
\par
As we discuss in the next section this is an example of Bianchi type
2A homogeneous cosmology with matter content: scalar matter.
\subsection{Elaboration of the solution with all roots switched on}
In \cite{piervoiastatia} we obtained a general solution of the
abstract $\mathrm{A_2}$ sigma--model where all the root
fields are switched on, generated  with the help of two
$\mathrm{SO(3)}$ rotations. It reads as follows.
\begin{eqnarray}
&& \nonumber
h_1(t) = \frac{t\,\left( -\kappa  + \omega  \right)  - 4\,\log (1 + e^{t\,\omega }) +
        2\,\log (1 + e^{t\,\omega } + e^{\frac{t\,\left( \kappa  + \omega  \right) }{2}})}{4\,{\sqrt{2}}}, \\
        && h_2(t) = \frac{t\,\left( \kappa  + 3\,\omega  \right)  -
        6\,\log (1 + e^{t\,\omega } + e^{\frac{t\,\left( \kappa  + \omega  \right) }{2}})}{4\,{\sqrt{6}}}, \\ &&
        \nonumber \varphi_1(t) = - \frac{1}{1 + e^{t\,\omega }} , \quad \varphi_2(t)   =    - \frac{1 + e^{t\,\omega }}{1 + e^{t\,\omega } + e^{\frac{t\,\left( \kappa  + \omega  \right) }{2}}} ,
          \quad \varphi_3(t) = \frac{1}{2\,\left( 1 + e^{t\,\omega } + e^{\frac{t\,\left( \kappa  +
             \omega  \right) }{2}} \right)}
             \label{tworotati}
\end{eqnarray}
By the procedure outlined above, in the canonical embedding of
$\mathrm{A_2}$, the solution (\ref{tworotati}) oxides to the
following Ricci-flat 5-dimensional metric:
\begin{eqnarray}
\nonumber && ds_5^2 = - e^{\ft 16 t(\kappa  + 3\omega
- 2\sqrt{3}\sqrt{\kappa^2 + 3\omega^2})}
          \left(1 + e^{t\omega} + e^{\frac{t(\kappa + \omega)}{2}} \right)dt^2 +  \\ && \nonumber
          + e^{\ft 16 t(\kappa  + 3\omega  - \sqrt{3}\sqrt{\kappa^2 + 3\omega^2})}
          \left(1 + e^{t\omega} + e^{\frac{t(\kappa + \omega)}{2}} \right)(\Omega_2^2 + \Omega_3^2)
          + \frac{e^{-\ft 16 t(\kappa - 3\omega)}}{(1+e^{t\omega})}
\Omega_1^2 + \\ && + \frac{e^{\frac{t\kappa}{3}}((1+e^{t\omega})\Omega_4 -\Omega_1)^2}{(1
+e^{t\omega})(1+e^{t\omega} +
e^{\ft 12 t(\kappa+\omega)})}
\label{rotaduemetra}
\end{eqnarray}
where, for shorthand notation, we have introduced the following differential forms:
\begin{eqnarray}
&& \nonumber \Omega_2 = dx_1, \quad \Omega_3 = dx_2, \\ &&
\Omega_1 = dx_3 + \ft {\omega}2(x_2dx_1 - x_1dx_2), \\ &&
\nonumber \Omega_4 = dx_4 - \ft {\kappa-\omega}{4}(x_2dx_1 -
x_1dx_2)
\end{eqnarray}
which close the following algebra:
\begin{eqnarray}
&& \nonumber d\Omega_2 = 0, \\
&& \nonumber d\Omega_3 = 0, \\
&& \nonumber d\Omega_1 = \omega \, \, \Omega_3\wedge\Omega_2, \\
&& d\Omega_4 = -\ft {\kappa-\omega}2 \,\Omega_3\wedge\Omega_2
\label{allrotforms}
\end{eqnarray}
\subsubsection{Dimensional reduction on a circle $S^1$}
The above five--dimensional metric can be reduced \'a la Kaluza Klein
on a circle $S^1$ and it  produces a further example of a Bianchi
type 2A metric which satisfies Einstein equations in presence of two
kinds of matter, a scalar field and a vector field.
\par
 To this effect we proceed as follows. We change the basis
in the algebra (\ref{allrotforms})
 \begin{equation}
 \bar{\Omega}_4 = \ft {(\kappa - \omega)}2 \,\Omega_1 + \omega \,\Omega_4 = \ft {(\kappa - \omega)}2 dx_3 + \omega \,
 dx_4 =dw,
 \quad d\bar{\Omega}_4 = 0
 \end{equation}
and we see  that the algebra we have obtained is essentially $Heis\times \mathbb{R}$.
Then the $5$-dimensional metric (\ref{rotaduemetra}) becomes:
\begin{eqnarray}
\nonumber && ds_5^2 = - e^{-\ft 16 t(\kappa  + 3\omega  - 2\sqrt{3}\sqrt{\kappa^2 + 3\omega^2})}
        \left(1 + e^{t\omega} + e^{\frac{t(\kappa + \omega)}{2}} \right)dt^2 +  \\ && \nonumber
        + e^{-\ft 16 t(\kappa  + 3\omega  - \sqrt{3}\sqrt{\kappa^2 + 3\omega^2})}
        \left(1 + e^{t\omega} + e^{\frac{t(\kappa + \omega)}{2}} \right)(\Omega_1^2 + \Omega^2_2) + \\ && \nonumber
      + \frac{e^{-\frac{t\kappa}{6}}[(\kappa-\omega)^2e^{\frac{t\kappa}{2}+t\omega}+
        4\omega^2e^{\frac{t\omega}{2}}+(\kappa+\omega)^2e^{\frac{t\kappa}{2}}]}
        {4\omega^2(1+e^{t\omega}+e^{\frac{1}{2}t(\kappa+\omega)})}\Omega_3^2 -\frac{e^{\frac{t\kappa}{3}}
        [(\kappa-\omega)(1+e^{t\omega})+2\omega]}{\omega^2(1+e^{t\omega}+e^{\frac{1}{2}t(\kappa+\omega)})}\Omega_3\otimes dw +
 \\ && + \frac{e^{\frac{t\kappa}{3}}(1+e^{t\omega})}{\omega^2(1+e^{t\omega}+e^{\frac{1}{2}t(\kappa+\omega)})}dw^2
\label{ds5corr}
\end{eqnarray}
and we can perform the dimensional reduction on the circle
parametrized by $w$. The result is easily obtained The dilaton is:
 \begin{equation}
 \phi = - \sqrt{\ft 3 2} \,\ft 12 \, \log[g_{ww}]  =  - \sqrt{\ft 3 2} \,\ft 12 \, \log \left[\frac{e^{\frac{t\,\kappa }{3}}\,\left( 1 + e^{t\,\omega } \right) }
          {\left( 1 + e^{t\,\omega } + e^{\frac{t\,\left( \kappa  + \omega  \right) }{2}} \right) \,{\omega }^2}\right]
 \label{dilatonpino}
 \end{equation}
The 4-dimensional metric in the Einstein frame reads as follows:
\begin{eqnarray}
  && \nonumber ds_{4E}^2 =   -\ft 1{\omega} e^{-\ft 12 t(\omega -2\sqrt{\frac{\kappa^2}{3} +
     \omega^2})}(1+e^{t\omega})^{\ft 12}
  (1+e^{t\omega}+e^{\ft 12 t(\kappa+\omega)})^{\ft 12}dt^2 + \\ && + \ft 1{\omega}
  e^{-\ft 12 t(\omega - \sqrt{\frac{\kappa^2}{3} + \omega^2})}(1+e^{t\omega})^{\ft 12}
  (1+e^{t\omega}+e^{\ft 12 t(\kappa+\omega)})^{\ft 12}(\Omega_2^2 + \Omega^3_2) + \\ \nonumber && +
  \ft 1{\omega}e^{\frac{t\omega}{2}}(1+e^{t\omega})^{-\ft 12}
  (1+e^{t\omega} + e^{\ft 12 t(\kappa + \omega)})^{-\ft 12}\Omega_1^2
  \label{ds4simp}
  \end{eqnarray}
 Due to the cross term $d\Omega_1\otimes dw$ in (\ref{ds5corr}), through dimensional reduction we obtain also a
 vector field, which is defined as:
 \begin{equation}
 g_{\mu w} = \gamma_{\mu}^{w}g_{ww},
 \quad \mu = 0,\cdots,3 \ee
 \be \gamma^w = -\ft 12 \left( \kappa  - \omega\tanh(\ft {t\omega}2) \right)(dx_3 + \ft {\omega}2(x_2dx_1 - x_1dx_2))
 \label{KKvec}
 \end{equation}
The fields (\ref{dilatonpino}, \ref{ds4simp}, \ref{KKvec}) are an
exact classical solution of the 4-dimensional $0$--brane action
(\ref{1-branad4}) with parameter ($a=- \sqrt{6}$) which is just
the dimensional reduction of the pure gravity lagrangian in five
dimensions.
\section{Cosmological metrics of Bianchi type 2A with $\mathrm{SO(2)}$ isotropy}
\label{bianchi} In this section we present a study of Bianchi
cosmological metrics of a particular choice, \textit{type 2A}. The
reason is that the $\mathrm{A_2}$ solutions of the
three--dimensional $E_{8(8)}/\mathrm{SO(16)}$ sigma model, once
oxided to the canonical representative of their Weyl orbit provide
exact gravity solutions precisely of this Bianchi type as we have
shown in the previous section.
\par
In the Bianchi classification of spatially homogeneous space--times,
which is a classification of three--dimensional algebras, type 2A
corresponds to a \textit{Heisenberg algebra} described by the Maurer Cartan
equations (\ref{MaurerCartan}).
An explicit realization of the differential algebra
(\ref{MaurerCartan}) is already suggested  by the results of the
previous section. In terms of cartesian coordinates   $x,y, z$ we have:
\begin{eqnarray} &&  \nonumber  {\Omega}_1  = - dz - \frac{\varpi
  }{4}\left( x\,dy - y\,dx \right), \\ && \nonumber  {\Omega}_2 =  dx,\\ &&
      {\Omega}_3  = dy
      \label{3dinvformsbis}
  \end{eqnarray}
The $1$--forms (\ref{3dinvformsbis}) are realized on
the group manifold obtained through the exponentiation of the
Heisenberg Lie algebra (\ref{Tvecti}) :
\begin{equation}
  \mathcal{G}_{Heis} \equiv \exp [Heis]
\label{gruppomanif}
\end{equation}
and the cartesian coordinates can be seen as parameters of such a
group. Occasionally, when convenient we can also use cylindrical
coordinates $(r\, , \, \theta \, , \, z)$ obtained via the
transformation
\begin{equation}
  x =r \, \cos \,  \theta \quad ; \quad y= r \, \sin \,  \theta
\label{xychange}
\end{equation}
As on any group manifold,  there exist on $\mathcal{G}_{Heis}$ two
mutually commuting sets of vector fields that separately satisfy the
Lie algebra of the group, the generators of the left translations and
the generators of the right translations. Let us agree that the
$1$--forms (\ref{3dinvformsbis}) are left invariant. Then the triplet of
vector fields that generate left translations $\overrightarrow{k}_i$
will be such that they satisfy the Lie algebra (\ref{Tvecti}) and
the Lie derivative of the $\Omega_i$ along them vanishes.
\begin{eqnarray}
 \left [ \overrightarrow{k}_i , \overrightarrow{k}_j \right ] & = & t_{ij}^\ell \, \overrightarrow{k}_\ell
 \label{killitre}\\
\ell_{\overrightarrow{k}_i} \Omega_j &= & 0 \label{invariaomeg}
\end{eqnarray}
The explicit form of such vector fields is the following one:
\begin{eqnarray}
\overrightarrow{k}_1 &=& \frac{\partial}{\partial z} \nonumber\\
  \overrightarrow{k}_2 & = &
\frac{\partial}{\partial x} - \frac{\varpi}{4} \, y \, \frac{\partial}{\partial z}\nonumber \\
\overrightarrow{k}_3 & = &
\frac{\partial}{\partial y} + \frac{\varpi}{4} \,x\, \frac{\partial}{\partial
z}\nonumber\\
\label{3dkillingCart}
\end{eqnarray}
The most general Bianchi type cosmological metric based on the
Heisenberg Lie algebra is obtained from (\ref{Bianchimet}) by
substituting the forms (\ref{3dinvformsbis}) and, for an arbitrary
choice of the time dependent matrix $h_{ij}(t)$, it admits the
vector fields (\ref{3dkillingCart}) as Killing vectors. The
resulting pseudo-Riemannian manifold is spatially homogeneous but
not isotropic. The space of metrics we want to consider is further
restricted by the requirement of an $\mathrm{SO(2)}$ isotropy. To
this effect we consider the following vector field
\begin{equation}
  \overrightarrow{k}_0 = \frac{\partial}{\partial\theta}  =   -x \,
  \frac{\partial}{\partial y} + y \, \frac{\partial}{\partial x}
\label{o2roto}
\end{equation}
which commutes with the vector fields (\ref{3dkillingCart})
and acts on the Maurer Cartan forms in the following way:
\begin{equation}
  \ell_{\overrightarrow{k}_0} \Omega_1 = 0 \quad ; \quad \ell_{\overrightarrow{k}_0}
  \Omega_2 = -\Omega_3 \quad ; \quad \quad \ell_{\overrightarrow{k}_0}
  \Omega_3 = \Omega_2
\label{rotatus}
\end{equation}
It follows from the above equation that the $1$--forms $\Omega_i$
arrange into a singlet and into a doublet of the $\mathrm{SO(2)}$
group generated by $\overrightarrow{k}_0$. Hence this latter will
also be a Killing vector of the Bianchi metric (\ref{Bianchimet})
if the matrix $h_{ij}$ is invariant under these $\mathrm{SO(2)}$
rotations, namely if it is of the form:
\begin{equation}
  h_{ij}(t)= \left( \begin{array}{ccc}
        \Delta(t) & 0 & 0 \\
        0 & \Lambda(t) & 0 \\
        0 & 0 & \Lambda(t) \
  \end{array}\right)
\label{hijspec}
\end{equation}
In conclusion the metrics of the following type, containing two
essential scale factors $\Lambda(t)$, $\Delta(t)$ \footnote{The
scale factor $A(t)$ can always be eliminated by a redefinition of
the time coordinate $t$}
\begin{equation}
  ds_{4}^2 = -A(t) \, dt^2 + \Lambda(t) \left( \Omega_2^2 + \Omega_3^2\right)
  +\Delta(t) \, \Omega_1^ 2
\label{metBd4gen}
\end{equation}
admit a four dimensional group of isometries:
\begin{equation}
  G = \mathcal{G}_{Heis} \, \times \, \mathrm{SO(2)}
\label{4paraiso}
\end{equation}
and the constant time sections of these space--times are
$3$--dimensional homogeneous spaces, with an $\mathrm{SO(2)}$
isotropy subgroup at each point.
\par
In euclidean coordinates the explicit form of the metric (\ref{metBd4gen}) reads as follows:
\begin{equation}
  ds^2_{4} = -A(t) \, dt^2 + \Lambda(t) \, \left( dx^2 + dy^2\right)
  + \Delta(t) \left( dz + \frac{\varpi}{4} (x \, dy - y \, dx) \right)
  ^2
\label{cartasmet}
\end{equation}
which turns out to be very useful in our subsequent discussion of geodesics.
\subsection{Einstein equations for the $\mathcal{G}_{Heis} \times \mathrm{SO(2)}$ Bianchi metrics}
We study under which conditions the metric (\ref{metBd4gen}) is a
solution of the Einstein field equations. To this effect we use the
vielbein formalism and we write the vierbein as follows:
\begin{equation}
  e^0 = \sqrt{A(t)} \, dt \quad ; \quad e^1 = \sqrt{\Delta(t)}
  \, \Omega_1 \quad ; \quad e^{2,3} = \sqrt{\Lambda(t)} \, \Omega^{2,3}
\label{vielbeinomsp}
\end{equation}
We can immediately calculate the spin connection from the vanishing torsion equation:
\begin{equation}
  d \, e^A + \omega^{AB} \, \wedge \, e^C \, \eta_{BC} = 0
\label{torsion-eq}
\end{equation}
where for the flat metric we have used the mostly plus convention:
\begin{equation}
  \eta_{ab} = \mbox{diag} \left\{ -,+,+,+\right\}
\label{etamostplus}
\end{equation}
We obtain the following result for the spin connection
\begin{equation}
  \begin{array}{ccccccc}
        \omega^{01} & = & \frac{\dot{\Delta}}{2\sqrt{A} \,\Delta} \, e^1 & ; &
        \omega^{02} & = & \frac{\dot{\Lambda}}{2\sqrt{A} \,\Lambda} \, e^2 \\
        \omega^{03} & = & \frac{\dot{\Lambda}}{2\sqrt{A} \,\Lambda} \, e^3& ;
          &\omega^{12} & = & - \omega \frac{\dot{\Delta}}{4 \,\Lambda} \, e^3 \\
        \omega^{13} & = &  \omega \frac{\dot{\Delta}}{4 \,\Lambda} \, e^2& ; &
        \omega^{23} & = &  \omega \frac{\dot{\Delta}}{4 \,\Lambda} \, e^1 \
  \end{array}
\label{spinconne}
\end{equation}
which can be used to calculate the curvature $2$--form and the Ricci
tensor from the standard formulae:
\begin{equation}
\begin{array}{cclcl}
  R^{AB} & \equiv & d\omega^{AB} + \omega^{AC} \, \wedge \, \omega^{DB} \, \eta_{CD} & = &
  R^{AB}_{\phantom{CD}CD} \, e^C \, \wedge \, e^D \\
  Ric_{FG} & = & \eta_{FA} \, R^{AB}_{\phantom{CD}GB} &  \null &
  \null
\end{array}
\label{ricciusgen}
\end{equation}
The  Ricci tensor turns out to be diagonal and has the following eigenvalues:
\begin{eqnarray}
Ric_{00} & = & \frac{A'(t)\, \Delta '(t)}{8\, {A(t)}^2\, \Delta (t)} + \frac{{\Delta \
'(t)}^2}{8\,
             A(t)\, {\Delta (t)}^2} + \frac{A'(t)\, \Lambda '(t)}{4\, {A(t)}^2\, \
\Lambda (t)} + \frac{{\Lambda '(t)}^2}{4\,
             A(t)\, {\Lambda (t)}^2}\nonumber\\
             && - \frac{\Delta ''(t)}{4\,
             A(t)\, \Delta (t)} - \frac{\Lambda ''(t)}{2\, A(t)\, \Lambda (t)} \nonumber\\
Ric_{11} & = & \frac{{\varpi }^2\, \Delta (t)}{16\, {\Lambda (t)}^2} - \frac{A'(t)\, \Delta \
'(t)}{8\, {A(t)}^2\, \Delta (t)} - \frac{{\Delta '(t)}^2}{8\,
             A(t)\, {\Delta (t)}^2} + \frac{\Delta '(t)\, \Lambda '(t)}{4\,
             A(t)\, \Delta (t)\, \Lambda (t)}\nonumber\\
             && + \frac{\Delta ''(t)}{4\,
             A(t)\, \Delta (t)}          \nonumber\\
Ric_{22} & = & Ric_{33}  \nonumber\\
Ric_{33} & = & -\frac{ {\varpi }^2\, \Delta (t)}{16\, {\Lambda
(t)}^2} - \ \frac{A'(t)\, \Lambda '(t)}{8\, {A(t)}^2\, \Lambda
(t)} + \frac{\Delta '(t)\, \ \Lambda '(t)}{8\, A(t)\, \Delta (t)\,
\Lambda (t)} + \frac{\Lambda ''(t)}{4\,
             A(t)\, \Lambda (t)}
\label{ricciusom}
\end{eqnarray}
With little more effort we can calculate the Einstein tensor  defined
by:
\begin{eqnarray}
  G_{AB} &=&Ric_{AB} -\ft 1 2 \, \eta_{AB} \, R \nonumber\\
  R&=& \eta^{FG} \, Ric_{FG}
\label{Einsteinusgen}
\end{eqnarray}
and we obtain a diagonal tensor with the following eigenvalues:
\begin{eqnarray}
G_{00} & = & -\frac{ {\varpi }^2\, \Delta (t) }{32\, {\Lambda
(t)}^2} + \ \frac{\Delta '(t)\, \Lambda '(t)}{4\,
             A(t)\, \Delta (t)\, \Lambda (t)} + \frac{{\Lambda '(t)}^2}{8\,
             A(t)\, {\Lambda (t)}^2} \nonumber\\
G_{11} & = & \frac{3\, {\varpi }^2\, \Delta (t)}{32\, {\Lambda (t)}^2} + \frac{A'(t)\, \
\Lambda '(t)}{4\, {A(t)}^2\, \Lambda (t)} + \frac{{\Lambda '(t)}^2}{8\,
             A(t)\, {\Lambda (t)}^2} - \frac{\Lambda ''(t)}{2\, A(t)\, \Lambda (t)} \nonumber\\
G_{22} & = & G_{33} \nonumber\\
G_{33} & = & -\frac{ {\varpi }^2\, \Delta (t)  }{32\, {\Lambda
(t)}^2} + \ \frac{A'(t)\, \Delta '(t)}{8\, {A(t)}^2\, \Delta (t)}
+ \frac{{\Delta \ '(t)}^2}{8\,
             A(t)\, {\Delta (t)}^2} + \frac{A'(t)\, \Lambda '(t)}{8\, {A(t)}^2\, \
\Lambda (t)}\nonumber\\
&& - \frac{\Delta '(t)\, \Lambda '(t)}{8\,
             A(t)\, \Delta (t)\, \Lambda (t)} + \frac{{\Lambda '(t)}^2}{8\,
             A(t)\, {\Lambda (t)}^2} - \frac{\Delta ''(t)}{4\,
             A(t)\, \Delta (t)} - \frac{\Lambda ''(t)}{4\, A(t)\, \Lambda (t)}
\label{Einsteinusspeom}
\end{eqnarray}
Let us now consider the matter contribution to the Einstein
equations for the above homogeneous but anisotropic universe. To
this effect we still need to consider the structure of the stress
energy tensor. The standard cosmological model is based on the use
of a perfect fluid description of matter namely, in curved index
notation, one writes \footnote{In the mostly minus conventions we
have   $ds^2 = g_{\mu \nu} dx^\mu \otimes dx^\nu$ and $g_{\mu
\nu}U^\mu U^\nu=-1$}:
\begin{equation}
  T^{\mu \nu } = \rho \, U^\mu \, U^\nu  + p \left( U^\mu \,
  U^\nu + g^{\mu \nu } \right)
\label{stressenten}
\end{equation}
where $\rho$ is the energy density, $p$ the pressure and $U^\mu$
the four-velocity field of the fluid. In isotropic and homogeneous
universes this fluid is assumed to be comoving. Namely,  the
velocity field is orthogonal to the \textit{constant time slices
of space--time} or equivalently it has vanishing scalar product
with all the six space--like Killing vectors:
\begin{equation}
  \left( \overrightarrow{U}, \overrightarrow{k}\right) = 0
\label{comovingfluid}
\end{equation}
In our chosen coordinate system this means $U=(1,0,0,0)$. More
intrinsically we can just state that in flat coordinates the stress
energy tensor has the following diagonal form:
\begin{equation}
  T_{AB} =\left( \begin{array}{cccc}
        \rho(t) & 0 & 0 & 0 \\
        0 & p(t) & 0 & 0 \\
        0 & 0 & p(t) & 0 \\
        0 & 0 & 0 & p(t) \
  \end{array}\right)
\label{stressopiatto}
\end{equation}
for the standard isotropic and homogeneous model.
\par
It is interesting that a very mild generalization of
eq.(\ref{stressopiatto}) can accommodate various models of matter,
arising from  a microscopic field theory representation. The generalization is just the following:
\begin{equation}
  T_{AB} =\left( \begin{array}{cccc}
        \rho(t) & 0 & 0 & 0 \\
        0 & \sigma(t) & 0 & 0 \\
        0 & 0 & p(t) & 0 \\
        0 & 0 & 0 & p(t) \
  \end{array}\right)
\label{stressopiatto2}
\end{equation}
where we have introduced two different pressure eigenvalues
$p_1(t) \equiv \sigma(t)$ and $p_{2,3}(t) =p(t)$ relative to the
the axis $1$ and $2,3$ respectively. The equality of the pressure
eigenvalues in the $2,3$ directions is just the consequence of the
$\mathrm{SO(2)}$ isotropy that we have assumed. It is very simple
and very useful to calculate the exterior covariant derivative of
the above tensor using the spin connection as determined in
eq.(\ref{spinconne}). We get:
\begin{eqnarray}
&\nabla T^{AB} = dT^{AB} + \omega^{AF} T^{GB} \,\eta_{FG} + \omega^{BF} T^{AF}
\,\eta_{FG}&\nonumber\\
&\null & \nonumber\\
& = \left(  \matrix{ \frac{{e^0}\,\rho '(t )}{{\sqrt{A(t )}}} & \frac{{e^1}\,
          \left( \rho (t ) + \sigma (t ) \right) \,\Delta '(t )}
          {2\,{\sqrt{A(t )}}\,\Delta (t )} & \frac{{e^2}\,
          \left( p(t ) + \rho (t ) \right) \,\Lambda '(t )}{2\,
          {\sqrt{A(t )}}\,\Lambda (t )} & \frac{{e^3}\,
          \left( p(t ) + \rho (t ) \right) \,\Lambda '(t )}{2\,
          {\sqrt{A(t )}}\,\Lambda (t )} \cr \frac{{e^1}\,
          \left( \rho (t ) + \sigma (t ) \right) \,\Delta '(t )}
          {2\,{\sqrt{A(t )}}\,\Delta (t )} & \frac{{e^0}\,
          \sigma '(t )}{{\sqrt{A(t )}}} & \frac{\varpi \,{e^3}\,
          {\sqrt{\Delta (t )}}\,\left( -p(t ) + \sigma (t ) \right) }
          {4\,\Lambda (t )} & \frac{\varpi \,{e^2}\,
          {\sqrt{\Delta (t )}}\,\left( p(t ) - \sigma (t ) \right) }
          {4\,\Lambda (t )} \cr \frac{{e^2}\,
          \left( p(t ) + \rho (t ) \right) \,\Lambda '(t )}{2\,
          {\sqrt{A(t )}}\,\Lambda (t )} & \frac{\varpi \,{e^3}\,
          {\sqrt{\Delta (t )}}\,\left( -p(t ) + \sigma (t ) \right) }
          {4\,\Lambda (t )} & \frac{{e^0}\,p'(t )}
      {{\sqrt{A(t )}}} & 0 \cr \frac{{e^3}\,
          \left( p(t ) + \rho (t ) \right) \,\Lambda '(t )}{2\,
          {\sqrt{A(t )}}\,\Lambda (t )} & \frac{\varpi \,{e^2}\,
          {\sqrt{\Delta (t )}}\,\left( p(t ) - \sigma (t ) \right) }
          {4\,\Lambda (t )} & 0 & \frac{{e^0}\,p'(t )}
      {{\sqrt{A(t )}}} \cr  }\right)& \nonumber\\
\label{extdertab}
\end{eqnarray}
Then we can easily calculate the divergence of the stress--energy tensor, obtaining:
\begin{eqnarray}
D_A\,T^{A0} & = & \frac{1}{2\sqrt{A(t)}}\left \{ \frac{\left[ \rho (t ) + \sigma (t ) \right] \,\Delta '(t )}
      {\Delta (t )} + \frac{2\,\left[ p(t ) + \rho (t ) \right] \,
          \Lambda '(t )}{\Lambda (t )} + 2\,\rho '(t)\,\right \}   \label{conserveq}\\
D_A\,T^{Ai} & = & 0 \quad ; \quad ( i= 1,\dots , 3)
\label{divergence}
\end{eqnarray}
Setting eq.(\ref{conserveq}) to zero is a conservation equation which
is necessary to impose and has to be satisfied in any consistent
solution.
\par
As a first example we can derive the equation of state of a free scalar field. It
suffices to calculate the stress energy tensor of such a field,
assuming that it depends only on time. Using the normalizations of
the action (\ref{dilagravact}), from the general formula:
\begin{equation}
  T_{\mu\nu }^{(scal)} = \ft 14 \left( \partial_\mu\phi \, \partial_\nu\phi  -\ft
  12 g_{\mu\nu} \, \partial_\rho\phi \, \partial_\sigma\phi \,
  g^{\rho\sigma}\right)
\label{genforsca}
\end{equation}
with a cosmological metric of type $ds^2 = g_{00} dt^2 + g_{ij} \, dx^i \,
dx^j$, we get:
\begin{equation}
  T_{00} = \ft 14 \, \dot{\phi}^2 \quad ; \quad T_{ij} = -\ft 14 g_{ij}
  g^{00} \,\dot{\phi}^2
\label{t00}
\end{equation}
Converting to flat indices and comparing with (\ref{stressopiatto2}) we identify the equation of state:
\begin{equation}
  \rho = - \ft 14 \, \dot{\phi}^2 \, g^{00} \quad ; \quad \sigma = p  = \rho
\label{persicone}
\end{equation}
Substituting such a relation into the conservation equation
(\ref{conserveq}) we obtain the following differential relation:
\begin{equation}
  \frac{\rho (t)\,\Delta '(\mu )}{{\sqrt{A(\mu )}}\,\Delta (\mu )} +
  \frac{2\,\rho (t)\,\Lambda '(\mu )}
      {{\sqrt{A(\mu )}}\,\Lambda (\mu )} + \rho '(t) = 0
\label{differela}
\end{equation}
which is immediately integrated to:
\begin{equation}
  \rho(t) = \frac{\mbox{const}}{\Lambda(t)^2 \, \Delta(t)}
\label{rhoconse}
\end{equation}
A second interesting example is provided by the case of a vector
gauge field  coupled to a dilaton. What we essentially consider is
the case of the $0$-brane action in four dimensions namely
eq.(\ref{1-branad4}), which leads to the Einstein equation
$G_{\mu\nu} = T_{\mu\nu}$ with the following stress energy
tensor:
\begin{equation}
  T_{\mu \nu }=\ft 1 4 \left (\partial_{\mu} \phi \partial_\nu \phi
  -\ft 12 \, g_{\mu \nu } \, \partial^\rho \phi \, \partial_\rho \phi
  \right)  - \ft 1 8 \, e^{-a \phi} \, \left ( F_{\mu \rho }\, F_{\nu \sigma
  } \, g^{\rho\sigma} - \ft 1 4 \, g_{\mu \nu } \, F_{\eta \rho }\, F_{\theta \sigma
  } \, g^{\rho\sigma} \, g^{\eta\theta}\right)
\label{branastresso}
\end{equation}
In the background of the metric (\ref{metBd4gen}) we introduce a
gauge $1$--form with the following structure:
\begin{equation}
  A = f(t) \, \Omega^1
\label{gauge1form}
\end{equation}
and we obtain the field strength $2$--form:
\begin{equation}
F =  \frac{\dot{f}(t)}{\sqrt{\Delta(t)}} \, A(t) \, e^0 \, \wedge
\, e^1 +  f(t) \, \frac{\varpi}{2 \Lambda(t)} \, e^2 \, \wedge \, e^3
\label{filastrenua}
\end{equation}
We can identify the intrinsic components of the $2$--form $F$ with
the electric and magnetic field as usual:
\begin{equation}
  F_{01}= \mathcal{E} = \ft 12 \frac{\dot{f}(t)}{\sqrt{\Delta(t) \,
  A(t)}}
  \quad ; \quad F_{23}= \mathcal{H} = \ft 1 4 \,  \varpi \, \frac{f(t) \,}{ \Lambda(t)}
\label{elecmagnet}
\end{equation}
and in terms of these items, the stress energy tensor
(\ref{branastresso}), reduced to flat indices, becomes of the form (\ref{stressopiatto2})
with:
\begin{eqnarray}
  \rho &=& \rho_{scal} + \rho_{vec}\nonumber\\
  \sigma &=& \rho_{scal} -\rho_{vec} \nonumber\\
 p &=& \rho_{scal} +\rho_{vec}
\label{tesnoron}
\end{eqnarray}
where
\begin{equation}
 \rho_{scal}=\ft 1
  4 \, \frac{\dot{\phi}(t)}{A(t)} \quad ; \quad \rho_{vec} = \ft 1 8 \,
  e^{-a \phi(t)} \left(\mathcal{E}^2(t)+\mathcal{H}^2(t)\right)
\label{rhovecrhosca}
\end{equation}
Given this setup we present three different exact solutions with
and without matter content. Once they are given  it is
straightforward to verify that they satisfy the coupled matter
equations, as we shall explicitly do, but it would be very
difficult to derive them in the context of General Relativity.
Indeed to our knowledge they had not been derived before, although
Bianchi classification is almost one century old. We obtained them
through the oxidation of the $\mathrm{A_2}$ solutions of the
three--dimensional sigma model as it was explicitly shown in previous
section.
\subsubsection{The vacuum solution and its properties}
It is a remarkable fact that we can obtain an exact solution of
the evolution equations in the absence of any matter content. What
we get is an empty Ricci flat universe with rather peculiar
properties. Imposing that the Ricci tensor (\ref{ricciusom})
vanishes (and hence also the Einstein tensor
(\ref{Einsteinusspeom})) we get differential equations for
$\Lambda(t)$, $\Delta(t)$ and $A(t)$ that are exactly solved by
the following choice of transcendental functions:
\begin{eqnarray}
A(t) & = & \exp(t\,\varpi) \, \cosh \left(\frac{t\,\varpi}{2}\right) \nonumber\\
\Lambda(t) & = & \exp(\frac{t\,\varpi}{2}) \, \cosh
\left(\frac{t\,\varpi}{2}\right)\nonumber\\
\Delta(t) & = &  \frac{1 }{\cosh\left(\frac{t\,\varpi}{2}\right)}
\label{factorila}
\end{eqnarray}
The scale factors (\ref{factorila}), which would be very hard to
determine by trying to solve Einstein equations directly, are
instead easily read off from the canonical oxidation of the
$\mathrm{A_2}$ model at $\kappa=0$, namely from
eq.(\ref{ds4}). It suffices to identify:
\begin{equation}
  \varpi = \omega
\label{varpi=omega}
\end{equation}
In order to write the metric in a standard cosmological form we need
to redefine the time variable by setting:
\begin{equation}
  \tau (t) = \int_{-\infty}^{t} \, \sqrt{A(t)} \, dt =  \int_{-\infty}^{t}
  e^{\frac{t\varpi}{2}}{\cosh^{\frac{1}{2}}(\frac{t\,\varpi }{2})}dt
  \label{choiceoftime}
\end{equation}

so that in the new cosmic time variable eq.(\ref{metBd4gen}) becomes:
\begin{equation}
  ds^2_{4}=   d\tau^2 + \Lambda(\tau) \left( \Omega_2^2 + \Omega_3^2\right)
  +\Delta(\tau) \, \Omega_1^2
\label{metBd4genBo}
\end{equation}
Equation (\ref{choiceoftime}) can be exactly integrated in terms of
hypergeometric functions. We obtain:
\begin{equation}
  \tau(t) = \frac{2\sqrt{2}}{3 \, \varpi} \, \exp \left[ \frac{t\,\varpi}{4}
  \right] \, \left( \sqrt{1+\exp[\tau \,\varpi]} + 2 \,
  \null_2F_1\left[\ft 14 ,\ft 12 , \ft 54 , -\exp[t\,\varpi] \right]
  \right )
\label{tautime}
\end{equation}
Although inverting eq.(\ref{tautime}) is not analytically
possible, yet it suffices to plot the behaviour of the scale
factors $\Lambda$ and $\Delta$ as functions of the cosmic time
$\tau$. This behaviour is shown in several graphics. In
fig.\ref{earlytime} we see the behaviour of the scale factors for
very early times.
\iffigs
\begin{figure}
\begin{center}
\epsfxsize =7cm
{\epsffile{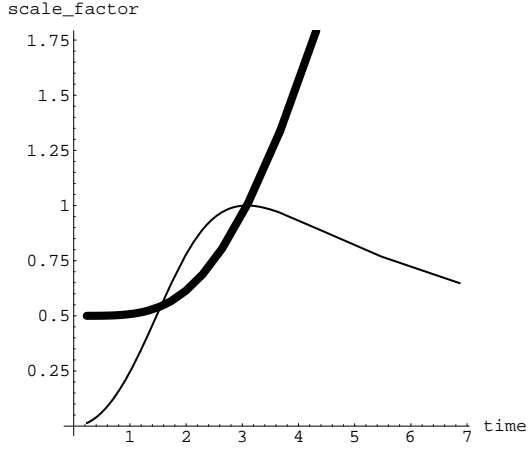}
\caption{\label{earlytime}Evolution of the cosmological scale
factors $\Lambda(\tau)$ (thick line) and $\Delta(\tau)$ (thin
line) for very early times, when the universe is very young for
the vacuum solution. $\Lambda$ starts at a finite value $0.5$ and
always grows, while $\Delta$ starts at zero, grows for some time
up to the maximum value $1$ and then starts decreasing } } \hskip
2cm \unitlength=1.1mm
\end{center}
\end{figure}
\fi
\par
The early finite behaviour of the scale factors has a very important
consequence. This space-time has no initial singularity. Indeed for
$\tau \mapsto 0$ the curvature $2$--form is perfectly well behaved
and tends to the following finite limit:
\begin{equation}
  \begin{array}{ccccccccccc}
        R^{01} & = & -\ft 1 2 E^2 \, \wedge \, E^3 & ; & R^{02} & = &- \ft 1 4  E^1 \,
        \wedge E^3 & ; &
        R^{03} & = & \ft 14 E^1 \, \wedge E^2 \\
        R^{12} & = & 0 & ; & R^{13} & = &
        0 & ; & R^{23} & = &
        \ft 12 E^0 \, \wedge E^1 \
  \end{array}
\label{curvat0}
\end{equation}
In fig.\ref{latedelta} we see the evolution of the $\Delta(\tau)$
and $\Lambda(\tau)$ scale factors for late times. Both of them
have a power-like asymptotic behaviour. $\Lambda(\tau)$ grows approximately as
 $\Lambda\sim\tau^\alpha$, $\alpha > 1 \quad (\approx \ft 32)$, and $\Delta(\tau)$ decreases as $\Delta\sim\tau^\beta$,
 $-1 < \beta < 0 \quad (\approx -\ft 7{10})$.

 \iffigs
\begin{figure}
\begin{center}
\epsfxsize =7cm {\epsffile{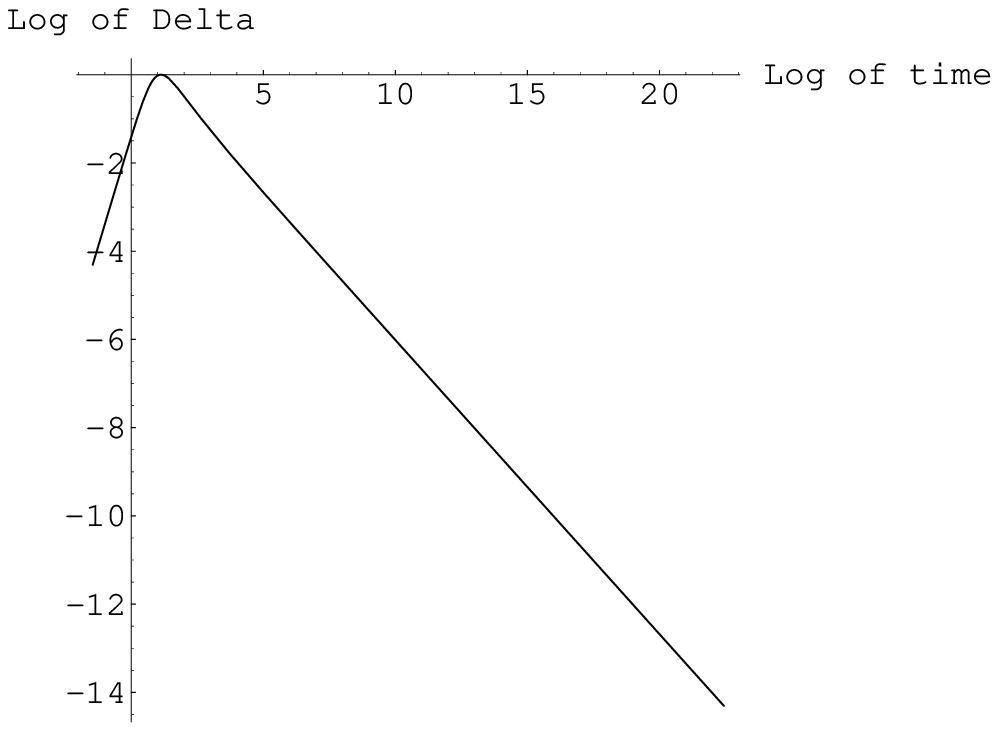}} \epsfxsize
=7cm {\epsffile{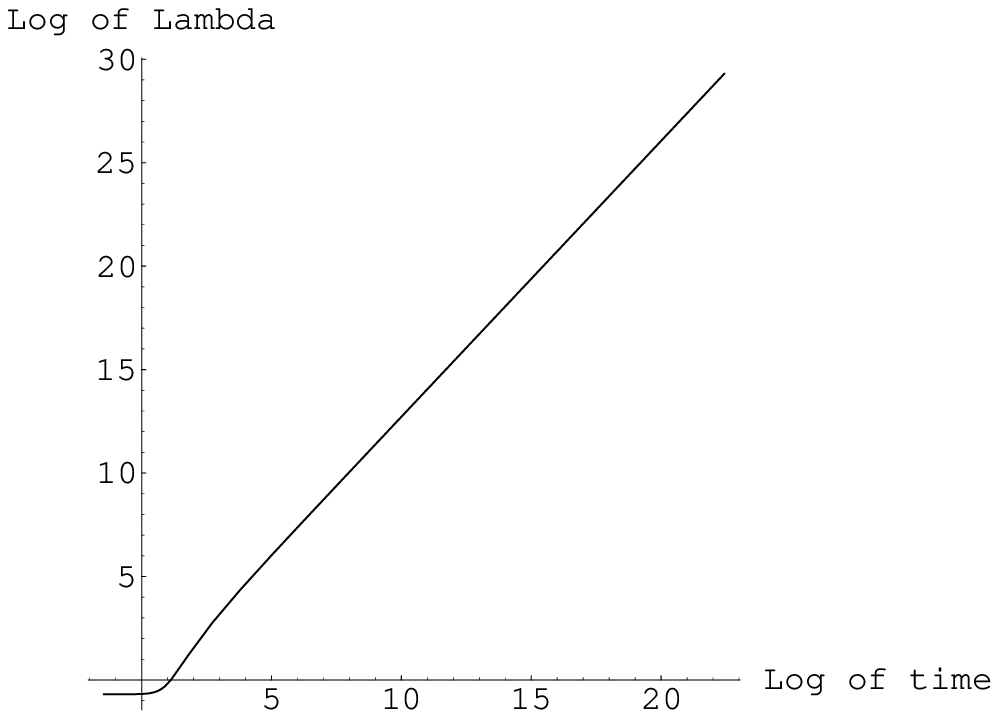}}
\caption{\label{latedelta}Evolution of the cosmological scale
factors $\Delta(\tau)$ and $\Lambda(\tau)$ for late times, the
graphic plots the logarithm of scale factor
  against the logarithm of cosmic time. $\Lambda$ continues to grow indefinitely in time with a power law.
  $\Delta$ tends to zero with
a power law.}
 \hskip 2cm \unitlength=1.1mm
\end{center}
\end{figure}
\fi
\par

We can summarize by saying that this funny homogeneous but not
isotropic universe, which is empty of matter, has a curious
history. It has no initial singularity but it is born finite,
small and essentially two--dimensional. It begins to expand and
the third dimension starts to develop. It reaches a state when it
is effectively three--dimensional, although still very small, the
two scale factors being of equal size. Then the third dimension
rapidly squeezes and the universe becomes again effectively two
dimensional growing monotonously large in the two dimensions in
which it was born.
\subsubsection{The dilaton gravity solution and its properties}
The second exact solution that we derive corresponds to a system
containing just a free dilaton field coupled to gravity. The
lagrangian is simply given by eq. (\ref{dilagravact}). If we
choose the following linear behaviour of the scalar field:
\begin{equation}
  \phi =  \ft 1 {2 \sqrt{6}} \kappa t
\label{scalarlinea}
\end{equation}
where $\kappa$ is some constant and we choose the following scale
factors,
\begin{eqnarray}
A(t) & = & e^{t\sqrt{\frac{\kappa^2}{3}+\varpi^2}}\cosh\frac{t\varpi}{2} \nonumber\\
\Lambda(t) & = & e^{\ft 12
t\sqrt{\frac{\kappa^2}{3}+\varpi^2}} \cosh\frac{t\varpi}{2}\nonumber\\
\Delta(t) &=&\frac{1}{\cosh\frac{t\varpi}{2}}
\label{dilagravscal}
\end{eqnarray}
by inserting into eq.(\ref{persicone}) we obtain:
\begin{equation}
  \rho = \frac{\kappa^2}{96} \, \frac{1}{A(t)} = \frac{\kappa^2}{96}
  \, e^{-t\sqrt{\frac{\kappa^2}{3}+\varpi^2}}\,
  \mbox{sech}\frac{t\varpi}{2} \quad ; \quad \sigma(t) = p(t) =  \rho(t)
\label{rhoscalavero}
\end{equation}
Comparison with eq.(\ref{rhoconse}) shows that indeed the energy
density in (\ref{rhoscalavero}) is of the required form and obeys the
conservation law, i.e. the field equation of the scalar field.  On
the other hand calculating the Einstein tensor, namely substituting
eq.s (\ref{dilagravscal}) into (\ref{Einsteinusspeom}) we get:
\begin{equation}
  G_{00}=G_{11}=G_{22}=G_{33} =\frac{\kappa^2}{96}
  \, e^{-t\sqrt{\frac{\kappa^2}{3}+\varpi^2}}\,
  \mbox{sech}\frac{t\varpi}{2}
\label{exactscalGtens}
\end{equation}
and in this way we verify that the Einstein equations are indeed
satisfied.
\par
Once again the scale factors (\ref{dilagravscal}) and the linear
behavior (\ref{scalarlinea}) of the scalar field that would be
very difficult to determine by solving Einstein equations
directly, are easily read off from the canonical oxidation of the
$\mathrm{A_2}$ model, namely from equations
(\ref{dilametrapre}) and (\ref{linescalapre}). Also in this case
we identify
\begin{equation}
  \varpi = \omega
\label{varpi=omeg2}
\end{equation}
\par
We can now investigate the properties of this solution. First of all
we reduce it to the standard form (\ref{metBd4genBo}) as we did in the
previous case. The procedure is the same, but now the cosmic time
$\tau$ has a different analytic expression in terms of the original
parametric time $t$.
Indeed, substituting the new form  of the scale function $A(t)$ as given in
eq.(\ref{dilagravscal}) into eq.(\ref{choiceoftime}) we obtain the following definition of the cosmic time:

\begin{eqnarray}
&&
  \tau (t) = \int_{-\infty}^{t} \,\sqrt{A(t)} \, dt =  \int_{-\infty}^{t} e^{\ft 12 t\sqrt{\frac{\kappa^2}{3} +
  {\varpi }^2}}\cosh^{\ft 12}(\frac{t\varpi }{2})dt = \\\nonumber
&& = \frac{2 e^{\frac{t( -\varpi  + \sqrt{\frac{\kappa^2}{3} + \varpi^2})}{2}}
\sqrt{(1 + e^{t \varpi}) \mbox{sech}\frac{t\varpi}{2}}}{-\varpi  +
2\sqrt{\frac{\kappa^2}{3} + \varpi^2}}  {_2 F_{1}}\left [- \ft 14  +
             \frac{\sqrt{\frac{\kappa^2}{3} + \varpi^2}}{2\varpi},\,
          - \ft 12 ,
          \ft 34 +  \frac{\sqrt{\frac{\kappa^2}{3} + \varpi^2}}{2\varpi},-e^{t\varpi } \right]
\label{lulla}
\end{eqnarray}
A plot of the function $\tau(t)$ for various values of $\kappa$ (see fig. \ref{cosmictime2}) shows
that $\tau$ has always the same qualitative behaviour. It tends
to zero for $t \mapsto -\infty$ and it grows exponentially for $t
\mapsto \infty$.
\iffigs
\begin{figure}
\begin{center}
\epsfxsize =7cm
{\epsffile{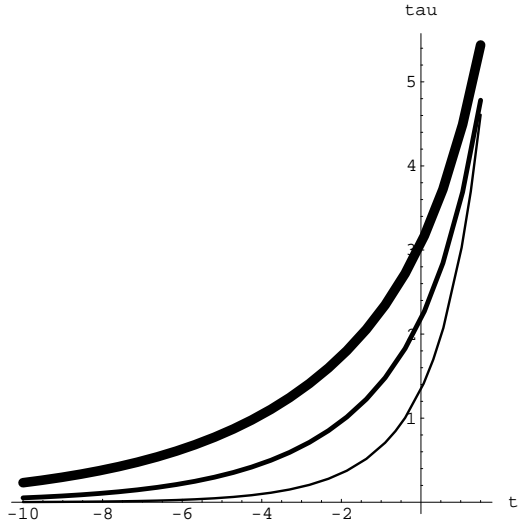}
\caption{\label{cosmictime2}The cosmic time $\tau$ versus the
parameter $t$  for various values of the parameter $\kappa$. The bigger $\kappa$ the thinner the corresponding line.
Here $\kappa=0$ is the thickest line. The other two correspond to $\kappa =1,2$ respectively.}
}
\hskip 2cm
\unitlength=1.1mm
\end{center}
\end{figure}
\fi

Hence we conclude that there is an initial time of this universe at
$\tau=0$ and we can explore the initial conditions.  In a completely
different way from the previous vacuum solution, this universe
displays an initial singularity and has a standard big bang
behaviour. The singularity can be seen in two ways. We can plot the
energy density  as given in eq.(\ref{rhoscalavero}) and realize that
for all values of $\kappa \ne 0$ it diverges at the origin of time
(see fig. \ref{enescal}).
\iffigs
\begin{figure}
\begin{center}
\epsfxsize =7cm
{\epsffile{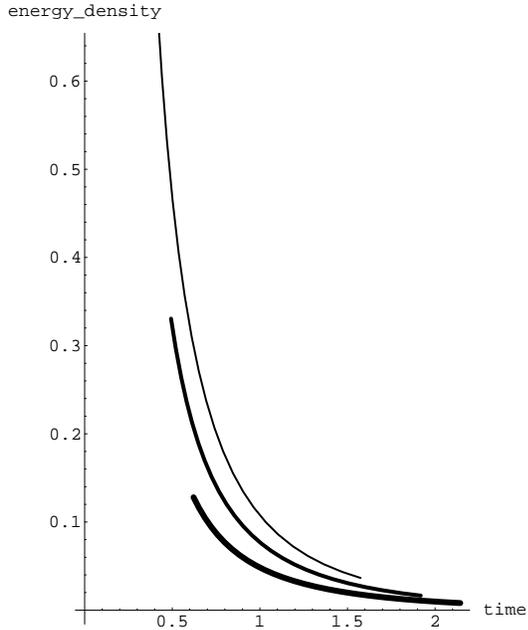}
\caption{\label{enescal} The evolution  of the energy density of the scalar field
as function of the cosmic time, for various values of $\kappa$. The bigger $\kappa$,
the thinner the corresponding line.
Here $\kappa=0.5$ is the thickest line. The other two correspond to $\kappa =0.7$ and
$\kappa=1$, respectively.}
}
\hskip 2cm
\unitlength=1.1mm
\end{center}
\end{figure}
\fi
\par
Alternatively, substituting the scale functions in the expression for
the curvature $2$--form, we can calculate its limit for $t\mapsto
-\infty$ and we find that the intrinsic components diverge for all
non vanishing values of $\kappa$, while they are finite at $\kappa=0$
as we have already remarked.
\par
Let us now analyze the behaviour of the two scale factors
$\Lambda(\tau)$ and $\Delta(\tau)$.
\iffigs
\begin{figure}
\begin{center}
\epsfxsize =7cm
{\epsffile{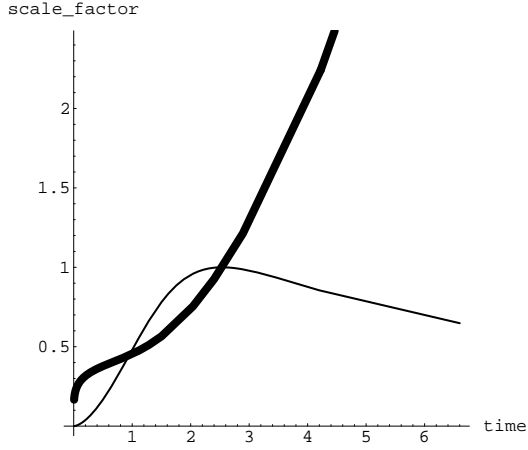}
\caption{\label{lamdelsca} The evolution  of the two scale factors as
function of the cosmic time $\tau$ in the dilaton gravity solution.
The thicker line is $\Lambda$ while
the thinner one is $\Delta$. The chosen value of the parameter kappa is $\kappa=0.7$.}
}
\hskip 2cm
\unitlength=1.1mm
\end{center}
\end{figure}
\fi
\par
This is displayed in fig.\ref{lamdelsca}. For late and intermediate
times the behaviour is just the same as in the vacuum solution with
$\kappa=0$, but the novelty is the behaviour of $\Lambda$ at the
initial time. Rather than starting from a finite value as in the
vacuum solution $\Lambda$ starts at zero just as $\Delta$. This is
the cause of the initial singularity and the standard big bang
behaviour.
Further insight in the behavior of this solution is obtained by
considering the evolution plots of the scale factors $\Lambda(\tau)$ and $\Delta(\tau)$
for various values of $\kappa$, see fig.(\ref{manylam1}).
\iffigs
\begin{figure}
\begin{center}
\epsfxsize =7cm
{\epsffile{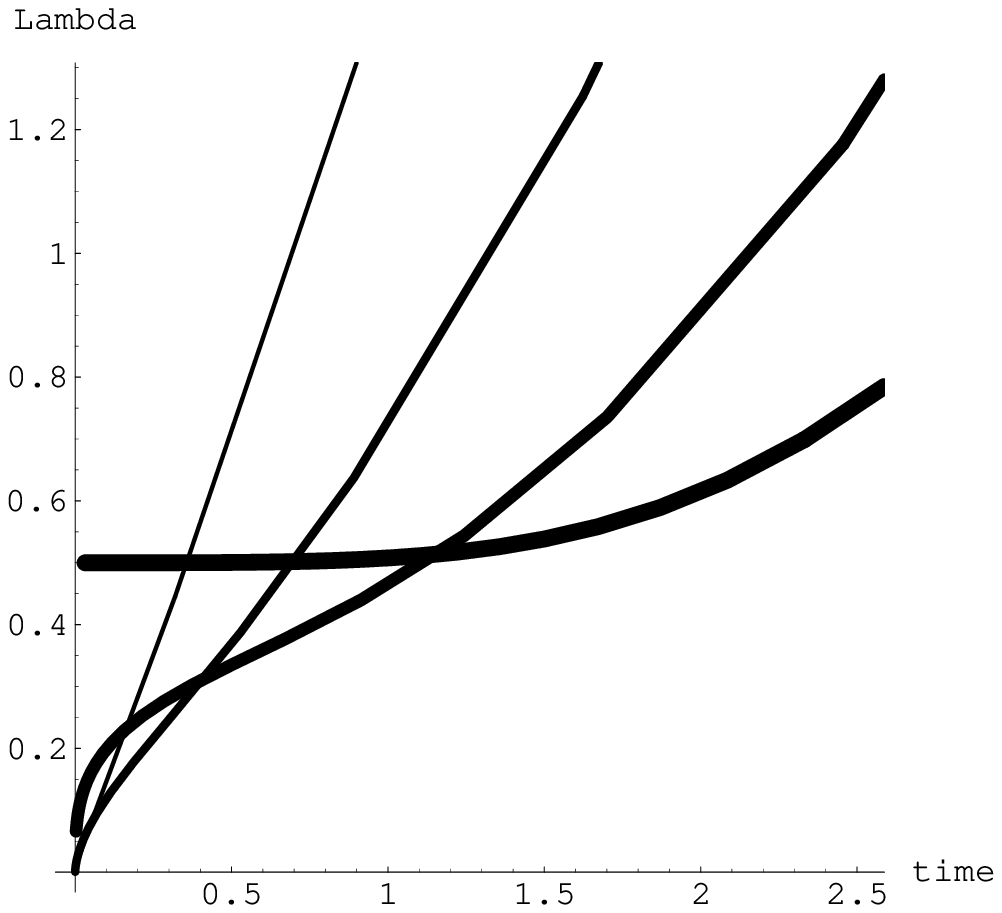}}
\epsfxsize =7cm
{\epsffile{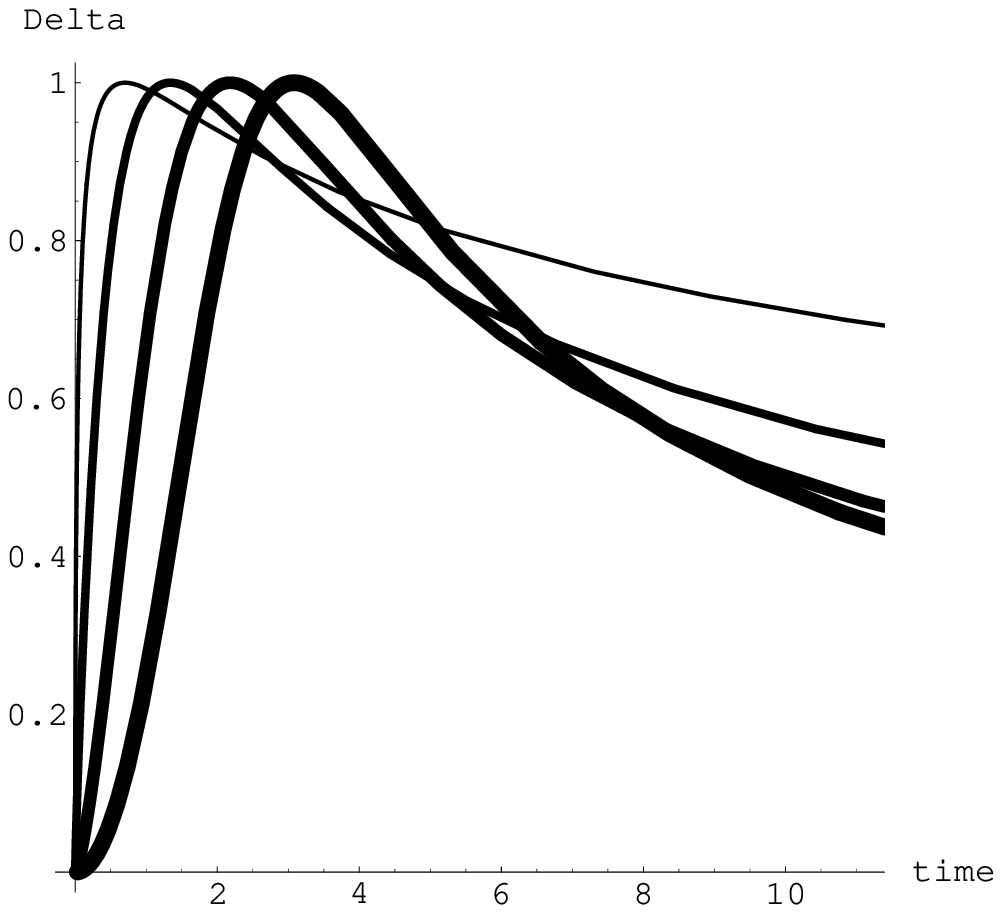}}
\caption{\label{manylam1} The evolution  of the scale factors $\Lambda$  and $\Delta$ as
functions of the cosmic time $\tau$ in the dilaton gravity solution and
for different values of kappa.
The thickest line corresponds to $\kappa=0$. The bigger $\kappa$,
the thinner the line. Here we have $\kappa=0,1,2,4$. For all $\kappa \ne 0$, $\Lambda$ begins at zero.
Instead $\Delta$ has always the same
behaviour and increasing $\kappa$ corresponds only to an anticipation of the peak.}
\hskip 2cm
\unitlength=1.1mm
\end{center}
\end{figure}
\fi

\subsubsection{The $0$-brane solution and its properties}
The third exact solution that we consider corresponds to the
$0$--brane system described by the lagrangian of eq.
(\ref{1-branad4}). Just as in the previous cases it would be very
difficult to integrate Einstein equations directly. Yet we can
read off an exact solution from our oxidation results in the
canonical embedding of the $\mathrm{A_2}$ model. It suffices
to set:
\begin{equation}
  \omega = \frac{\varpi}{2}
\label{omega=varpimez}
\end{equation}
and from eq.s (\ref{ds4simp}), (\ref{KKvec}) and (\ref{dilatonpino})
we immediately obtain the required data. So
the solution is obtained by choosing the following form for the
dilaton
\begin{equation}
\phi(t) =  -\frac{\sqrt{3}}{2\sqrt{2}}\log \left[\frac{4e^{\frac{t\kappa}{3}}
( 1 + e^{\frac{t\varpi}{2}} ) }{( 1 + e^{\frac{t \varpi }{2}} +
e^{\frac{t( 2\,\kappa  + \varpi) }{4}}){\varpi }^2} \right]
\label{branasolscal}
\end{equation}
the following form for the gauge field
\begin{equation}
  \gamma = - \ft 12\left(\kappa  - \frac{\varpi}{2}\tanh\frac{t\varpi}{4}\right) \Omega_1
\label{gaugefield}
\end{equation}
the following value for the parameter $a$:
\begin{equation}
  a=-\sqrt{6}
\label{aparam}
\end{equation}
and the following form for the scale factors:
\begin{eqnarray}
A(t) & = & \ft 2{\varpi} e^{- t(\frac{\varpi}{4} -\sqrt{\frac{\kappa^2}{3} + \frac{\varpi^2}{4}})}
(1+e^{\frac{t\varpi}{2}})^{\ft 12}(1+e^{\frac{t\varpi}{2}}+e^{\ft 14 t(2\kappa+\varpi)})^{\ft 12} \nonumber\\
                          \null &\null &\null \nonumber\\
\Lambda(t) & = & \ft 2{\varpi} e^{- t(\frac{\varpi}{4} -\frac{1}{2}\sqrt{\frac{\kappa^2}{3} + \frac{\varpi^2}{4}})}
(1+e^{\frac{t\varpi}{2}})^{\ft 12}(1+e^{\frac{t\varpi}{2}}+e^{\ft 14 t(2\kappa+\varpi)})^{\ft 12}\nonumber\\
                 \null &\null &\null \nonumber\\
\Delta(t) & = &\ft 2{\varpi}e^{\frac{t\varpi}{4}}(1+e^{\frac{t\varpi}{2}})^{-\ft 12}
  (1+e^{\frac{t\varpi}{2}} + e^{\ft 14 t(2\kappa+\varpi)})^{-\ft 12}
\label{baransoluzia}
\end{eqnarray}
Inserting these data into eq.s (\ref{tesnoron}) and
(\ref{rhovecrhosca}) for the stress energy tensor
and in the expression for the Einstein tensor (\ref{Einsteinusspeom})
we can explicitly verify that with these choices the field equations are indeed satisfied since
we have:
\begin{eqnarray}
\rho(t) &=& G_{00} = G_{22}=G_{33} \nonumber\\
\sigma(t) &=& G_{11}
\end{eqnarray}
where $\rho(t)$ and $\sigma(t)$ are two explicit functions of all the
parameters whose expression is too long and messy to be reported, but
which can be straightforwardly computed from their definitions. It is rather
more convenient to plot them. In order to do that we need first to
reduce the metric to the standard form (\ref{metBd4genBo}). This time
the integration of the function $\sqrt{A(t)}$ cannot be done
analytically and we must confine ourselves to define the numerical
function:
\begin{equation}
  \tau(t) = \int_{-\infty}^t \, dt \ft 2{\varpi} e^{- t(\frac{\varpi}{4} -\sqrt{\frac{\kappa^2}{3} + \frac{\varpi^2}{4}})}
(1+e^{\frac{t\varpi}{2}})^{\ft 12}(1+e^{\frac{t\varpi}{2}}+e^{\ft 14 t(2\kappa+\varpi)})^{\ft 12}
\label{perdifiato}
\end{equation}
Plotting the energy density and pressure in the direction of
$\Omega_1$ (see fig.\ref{eneveca}) we see that once
again in this model there is an initial singularity at $\tau=0$,
since for all values of $\kappa$ the energy density diverges as $\tau
\mapsto 0$.
\iffigs
\begin{figure}
\begin{center}
\epsfxsize =4.5cm
{\epsffile{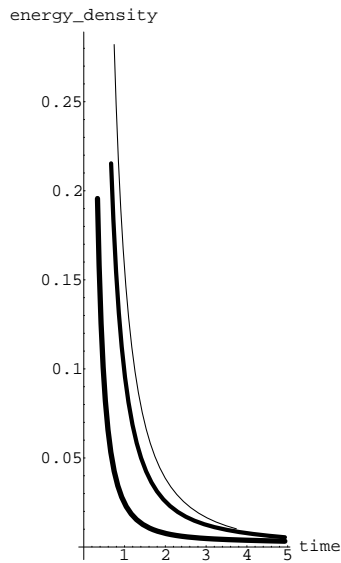}}
\epsfxsize =6cm
{\epsffile{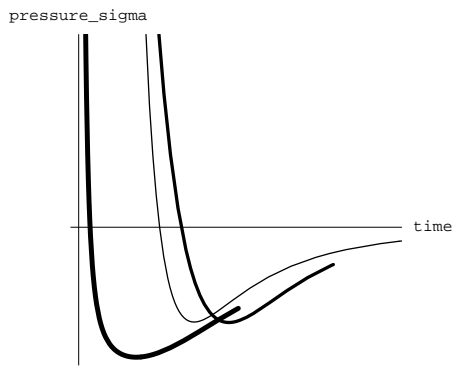}}
\caption{\label{eneveca}  Also in the $0$-brane solution as in the dilaton solution,
the energy density becomes infinitely large as $\tau \mapsto
0$, for all values of $\kappa$. The pressure eigenvalue in the
Cartan Maurer direction $\Omega_1$ suffers a minimum for all values of $\kappa$. This indicates that there should be
a bouncing phenomenon (billiard) in the corresponding scale factor
for which we expect a maximum. As in previous tables we distinguish the values
of $\kappa$ by the thickness of the line.
The thinner the line the larger $\kappa$.}
\unitlength=1.1mm
\end{center}
\end{figure}
\fi
\par
Considering now the plot of the pressure, we see that there is always a
minimum for all values of $\kappa$. This is the symptom that there
should be a billiard phenomenon in such a direction.

Indeed the plot of the $\Delta(\tau)$ scale factor is fully analogous
in shape to the previous cases and displays a peak with a maximum for
all values of $\kappa$ (see fig.\ref{manydeltvec})
\iffigs
\begin{figure}
\begin{center}
\epsfxsize =6cm
{\epsffile{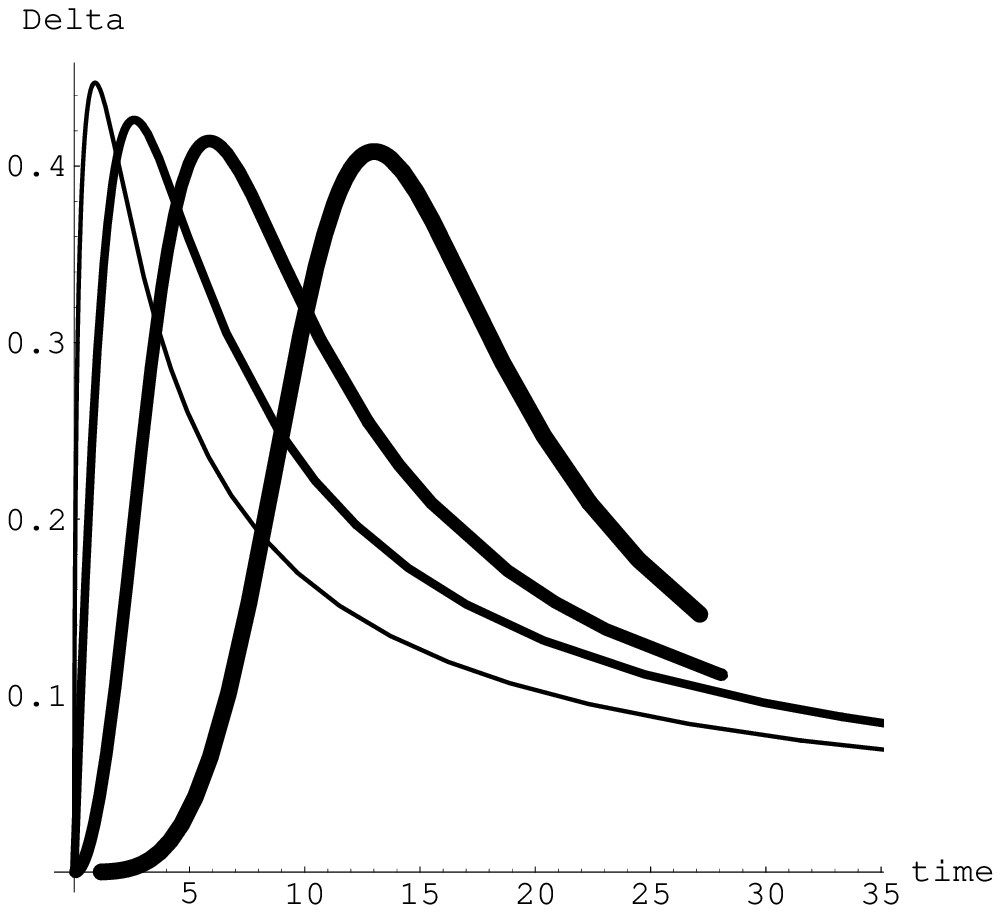}}
\epsfxsize =6cm
{\epsffile{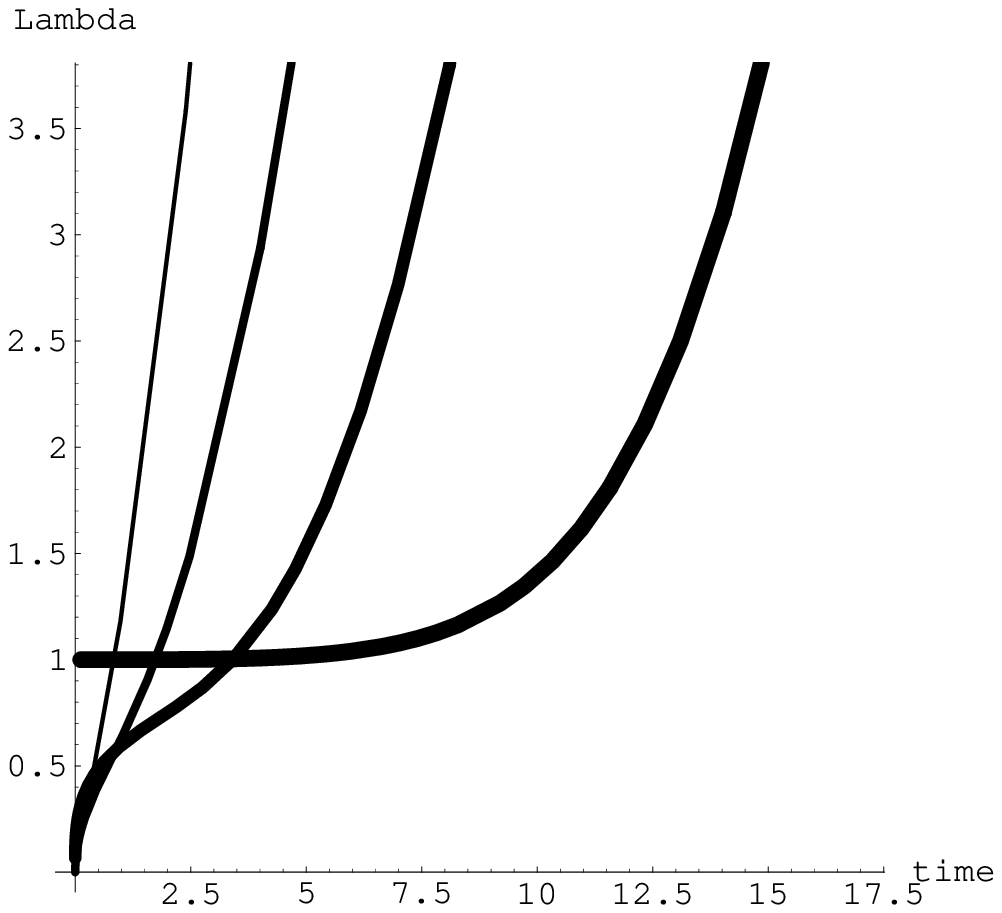}}
\caption{\label{manydeltvec}  In the $0$-brane solution  as in all
the other cases the $\Delta$ scale factor has a well pronounced
maximum for all values of $\kappa$. This is the billiard phenomenon.
The $\Lambda$ scale factor begins at zero for all non
vanishing values of $\kappa$ and it grows indefinitely. As usual thicker the line smaller the value of $\kappa$.}
\unitlength=1.1mm
\end{center}
\end{figure}
\fi
\par
The behaviour of the scale factor $\Lambda(\tau)$ is instead the same as
in the case of the dilaton gravity solution: see fig.\ref{manydeltvec}.

For $\kappa=0$ the scale factor $\Lambda(t)$ begins at a finite
value. Yet, differently from the case of the vacuum solution,
notwithstanding this fact the curvature $2$--form is singular in the
limit $\tau \mapsto 0$. This is consistent with the divergence of the
energy density and means that we have a standard big bang behaviour
for early enough times.
\subsection{Geometry of the homogeneous three-space and geodesics}
In order to better appreciate the structure of the
cosmological  solutions we have been considering in the previous subsection it is
convenient to study the geometry of the \textit{constant time
sections} and the shape of its geodesics. At every instant of time we
have the $3D$--metric:
\begin{equation}
  ds^2_{3} = \Lambda \, \left( dx^2 + dy^2 \right) + \Delta \left[
  dz + \frac{\varpi}{4}\left(xdy  -ydx\right) \right]^2
\label{3dmetra}
\end{equation}
which admits the Killing vectors (\ref{3dkillingCart}) as generators
of isometries. As it is well known, the scalar product of Killing
vectors with the tangent vector to a geodesic is constant along the
geodesic. Hence if $\lambda$ is the affine parameter along a
geodesic and $\overrightarrow{t}=\{x'[\lambda], y'[\lambda], z'[\lambda]\}$
is the tangent vector to the same, then we have the following four constants of
motion:
\begin{eqnarray}
A_1 \equiv (\overrightarrow{k}_1 \, , \, \overrightarrow{t}) & = &
\frac{\Delta \,\left( -\left( \varpi \,y(\lambda )\,x'(\lambda ) \right)  + \varpi \,x(\lambda )\,y'(\lambda ) +
             4\,z'(\lambda ) \right) }{4}\nonumber\\
A_2 \equiv (\overrightarrow{k}_0 \, , \, \overrightarrow{t})& = & \frac {1} {16} \, \left [\left( 16\,\Lambda
 + \Delta \,{\varpi }^2\,{x(\lambda )}^2 \right) \,y(\lambda )\,x'(\lambda ) +
  \Delta \,{\varpi }^2\,{y(\lambda )}^3\,x'(\lambda ) \right.\nonumber\\
  &&\left. -
  \Delta \,\varpi \,{y(\lambda )}^2\,\left( \varpi \,x(\lambda )\,y'(\lambda ) + 4\,z'(\lambda ) \right) \right.
  \nonumber\\
  &&\left. -
  x(\lambda )\,\left( \left( 16\,\Lambda  + \Delta \,{\varpi }^2\,{x(\lambda )}^2 \right) \,y'(\lambda ) +
          4\,\Delta \,\varpi \,x(\lambda )\,z'(\lambda ) \right) \right]\nonumber\\
A_3 \equiv (\overrightarrow{k}_2 \, , \, \overrightarrow{t})& = &
\frac{\left( 8\,\Lambda  + \Delta \,{\varpi }^2\,{y(\lambda )}^2 \right) \,x'(\lambda ) -
        \Delta \,\varpi \,y(\lambda )\,\left( \varpi \,x(\lambda )\,y'(\lambda ) + 4\,z'(\lambda ) \right) }{8} \nonumber\\
A_4 \equiv (\overrightarrow{k}_3 \, , \, \overrightarrow{t})& = & \frac{8\,\Lambda \,y'(\lambda )
 + \Delta \,{\varpi }^2\,{x(\lambda )}^2\,y'(\lambda ) +
        \Delta \,\varpi \,x(\lambda )\,\left( -\left( \varpi \,y(\lambda )\,x'(\lambda ) \right)
          + 4\,z'(\lambda ) \right) }{8}
\nonumber\\
\label{a1-a4}
\end{eqnarray}
Then the geodesics are characterized by the equations:
\begin{equation}
  A_2 = \frac{-4\,{A_4}\,x(\lambda )
      + 4\,{A_3}\,y(\lambda ) + \varpi \,{A_1}\,\left( {x(\lambda )}^2 + {y(\lambda )}^2 \right) }{4}
\label{equageodesa1}
\end{equation}
and
\begin{equation}
  {{z'(\lambda )}  =   {\frac{8\,\Delta \,\varpi \,{A_2} + {A_1}\,\left( 8\,\Lambda  - \Delta \,{\varpi }^2\,{x(\lambda )}^2 -
                      \Delta \,{\varpi }^2\,{y(\lambda )}^2 \right) }{8\,\Delta \,\Lambda }}}
\label{equageodesa2}
\end{equation}
We also have:
\begin{eqnarray}
 & & {{x'(\lambda )}= {\frac{2\,{A_3} + \varpi \,{A_1}\,y(\lambda )}{2\,\Lambda }}} \nonumber\\
 & & {{y'(\lambda )}= {\frac{2\,{A_4} - \varpi \,{A_1}\,x(\lambda )}{2\,\Lambda }}}
\label{alsohave}
\end{eqnarray}
We conclude that the projection of all geodesics on the $xy$ plane
are circles with centers at:
\begin{equation}
  (x_0 \, , \, y_0 ) = \left ( \frac{2\,{A_4}}{\varpi \,{A_1}} \, , \,\frac{-2\,{A_3}}{\varpi
  \,{A_1}}\right)
\label{centri}
\end{equation}
and radii:
\begin{equation}
  R = 2\,{\sqrt{\frac{\varpi \,{A_1}\,{A_2} + {{A_3}}^2 + {{A_4}}^2}{{\varpi }^2\,{{A_1}}^2}}}
\label{radiuszi}
\end{equation}
and in terms of the new geometrically identified constants eq.(\ref{equageodesa2}) becomes:
\begin{equation}
  {{z'(\lambda )}   =  {\frac{{A_1}\,\left( 8\,\Lambda  + 2\,\Delta \,{\varpi }^2\,\left( R^2 - {{x_0}}^2 - {{y_0}}^2 \right)  -
                     \Delta \,{\varpi }^2\,{x(\lambda )}^2 -
                      \Delta \,{\varpi }^2\,{y(\lambda )}^2 \right) }{8\,\Delta \,\Lambda }}}
\label{equageodesia3}
\end{equation}
If we use a polar coordinate system in the $xy$-plane, namely if we write:
\begin{eqnarray}
  x_0=\rho \, \cos \, [\theta] & ; &  y_0=\rho \, \sin \,
  [\theta] \nonumber\\
        x=\rho \, \cos \, [\theta] + R \, \cos[\varphi(\lambda)]& ; &
  x=\rho \, \sin \, [\theta] + R \, \sin[\varphi(\lambda)]
\label{polarplan}
\end{eqnarray}
where $\rho$ and $\theta$ are constant parameters,
we obtain that the derivative of the angle $\phi$ with respect to the affine parameter $\lambda$ is just:
\begin{equation}
 \frac{d\varphi}{d\lambda} = - \frac{\varpi \, A_1}{2 \, \Lambda}
\label{dfidlam}
\end{equation}
This means that $\varphi$ itself, being linearly related to $\lambda$, is an affine parameter.
On the other hand, the equation for the coordinate $z$,
(\ref{equageodesa2}), becomes:
\begin{equation}
\frac{dz}{d\varphi} =  \frac{-\left( 8\,\Lambda  +
             \Delta \,\left( R^2 - 3\,{\rho }^2 \right) \,
                {\varpi }^2 - 2\,R\,\Delta \,\rho \,{\varpi }^2\,
                \cos (\theta  - \varphi (\lambda )) \right) }{4\,
        \Delta \,\varpi }
\label{dzdfi}
\end{equation}
which is immediately integrated and yields:
\begin{equation}
  z[\varphi]=\frac{\left( \theta  - \varphi  \right) \,
          \left( 8\,\Lambda  +
                \Delta \,\left( R^2 - 3\,{\rho }^2 \right) \,
                 {\varpi }^2 \right)  -
        2\,R\,\Delta \,\rho \,{\varpi }^2\,
          \sin (\theta  - \varphi )}{4\,\Delta \,\varpi }
\label{zofpfi}
\end{equation}
Hence the possible geodesic curves in the three--dimensional sections
of the cosmological solutions we have been discussing  are described
by eq.(\ref{zofpfi}) plus the second of eq.s (\ref{polarplan}). The
family of such geodesics  is parametrized by $\{ R,\theta,\rho \}$,
namely by the position of the center in the $xy$ plane and by the
radius. The shape of such geodesics is that of spirals (see
fig.\ref{geodesia12}).
\iffigs
\begin{figure}
\begin{center}
\epsfxsize =5.5cm
{\epsffile{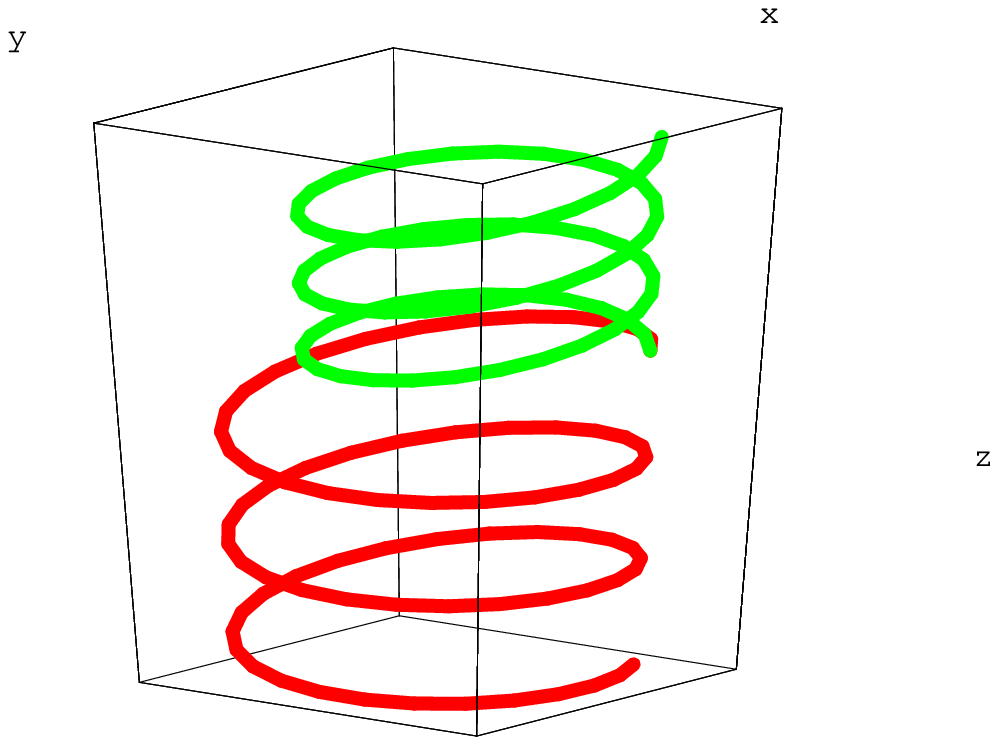}
\epsfxsize =5.5cm
\epsffile{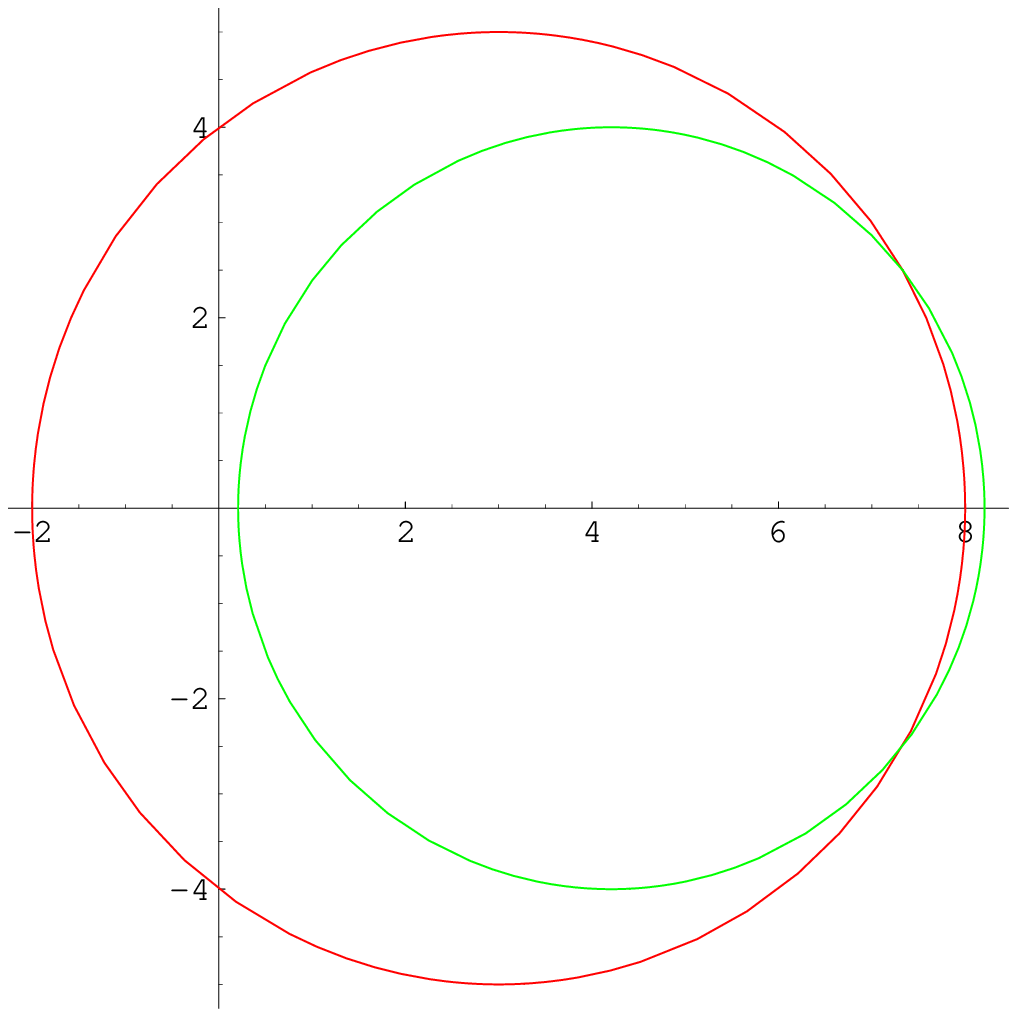}
\caption{\label{geodesia12}  In the first picture we see two geodesics
in three space, while in the second we see their projection onto the
plane $xy$.
 }
}
\unitlength=1.1mm
\end{center}
\end{figure}
\fi
\par
A more illuminating visualization of this three--dimensional geometry is
provided by the picture of a congruence of geodesics. Given a point in
this $3D$ space, we can consider all the geodesics that begin at
that point and that have a radius $R$ falling in some interval:
\begin{equation}
  R_A < R < R_B
\label{spanofR}
\end{equation}
Following each of them for some amount of  \textit{parametric time}
$\lambda$ we generate a two dimensional surface. An example is given
in fig.\ref{congruence1}.
\iffigs
\begin{figure}
\begin{center}
\epsfxsize =5.5cm
{\epsffile{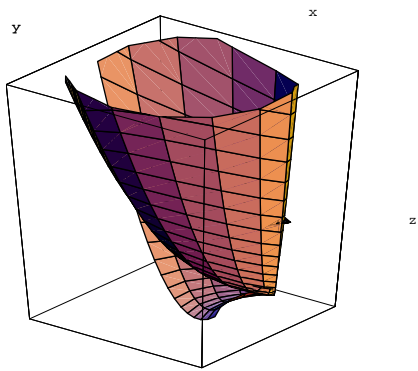}
\caption{\label{congruence1}  In this  picture we present a congruence of geodesics
for the space with $\Lambda=\Delta=\varpi=1$. All the curves start
from the same point and are distinguished by the value of the radius
$R$ in their circular projection onto the $xy$ plane.
 }
}
\hskip 2cm
\unitlength=1.1mm
\end{center}
\end{figure}
\fi
\par
The evolution of the universe can now be illustrated by its effect on
a congruence of geodesics.  Chosen a congruence like in
fig.\ref{congruence1}, the shape of the surface generated by such a  congruence depends
on the value of the scale parameters $\Lambda$ and $\Delta$. We can
follow the evolution of the congruence while the universe expands obtaining a movie. In
fig.\ref{congruomov} we present six photograms of such a movie:
\iffigs
\begin{figure}
\begin{center}
\vskip 1cm
{\epsfxsize =5cm
\epsffile{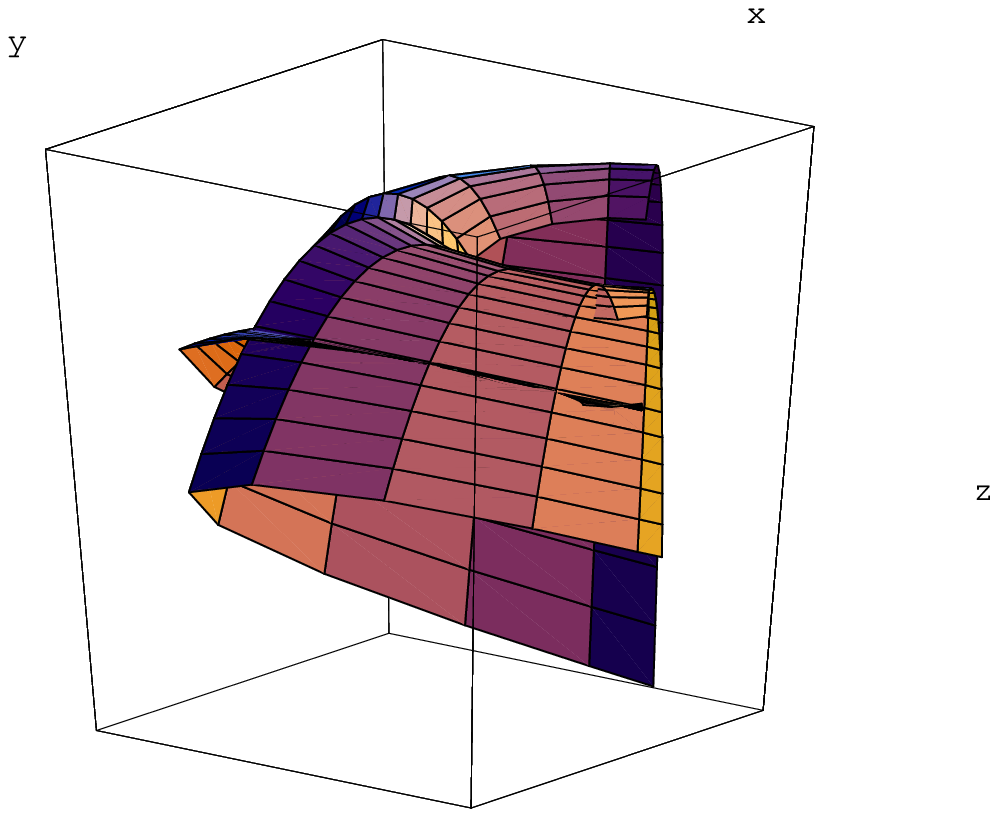}
\epsfxsize =5cm
\epsffile{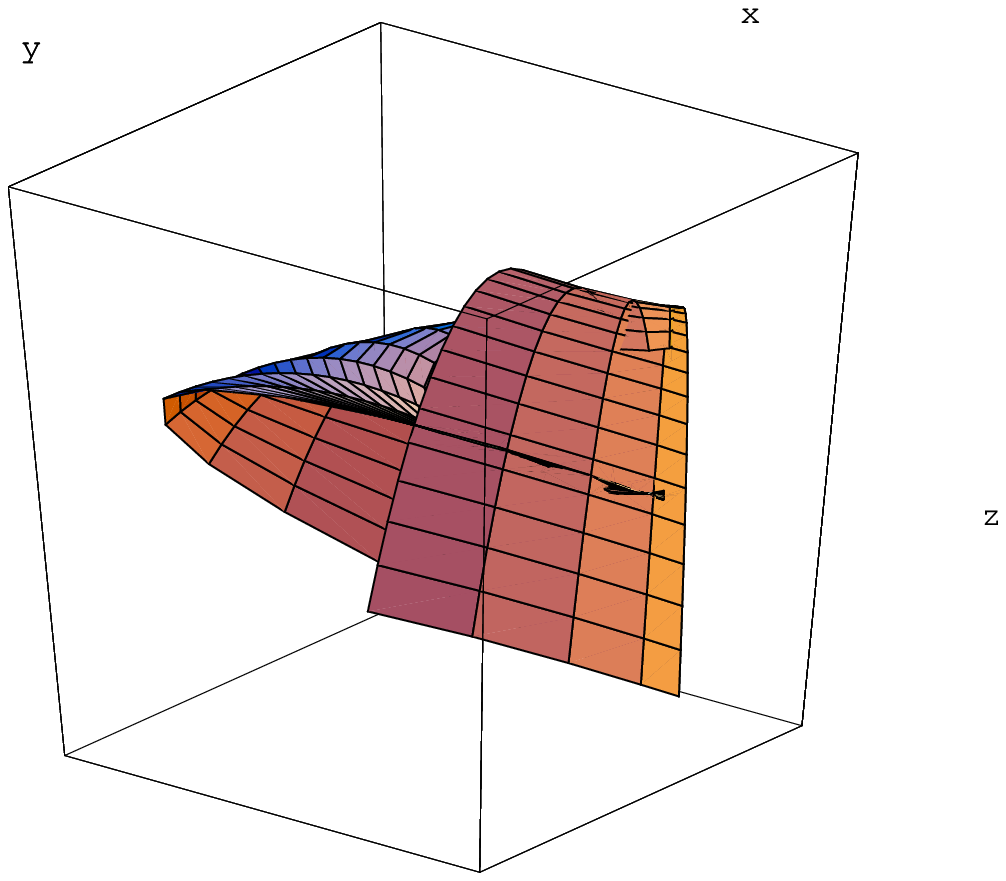}
\epsfxsize =5cm
\epsffile{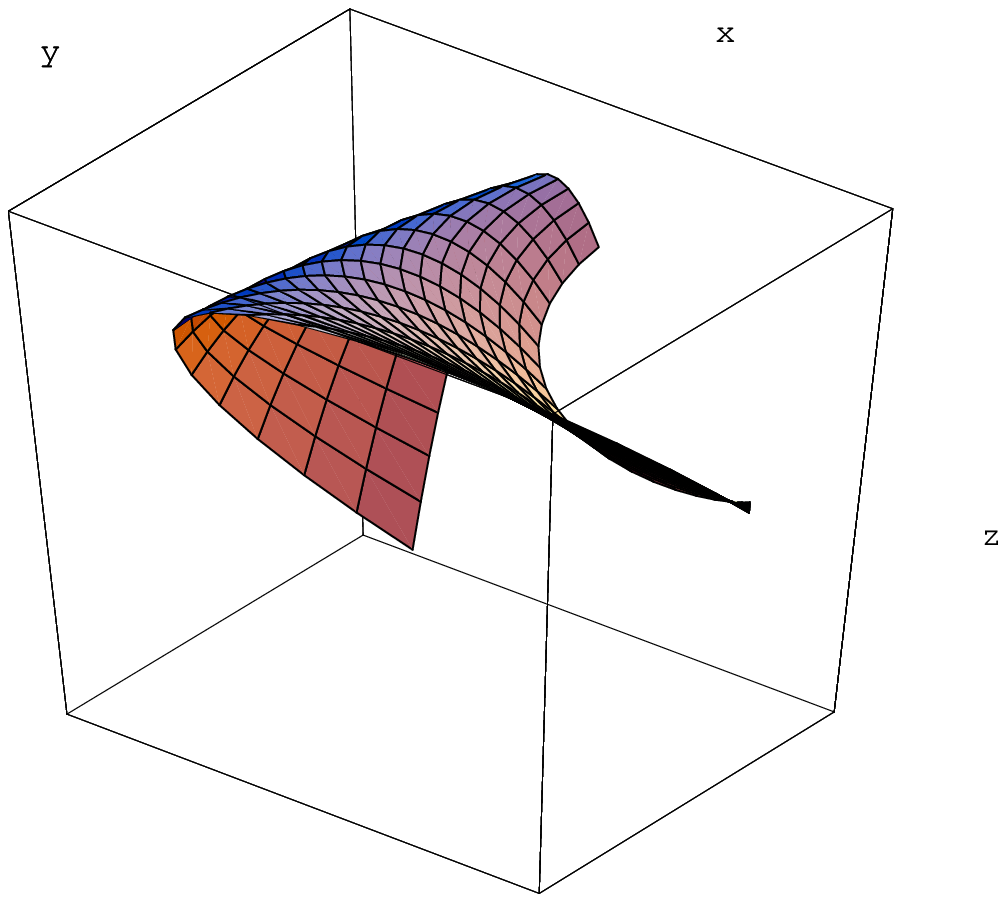}
\vskip0.5cm
\epsfxsize =5cm
\epsffile{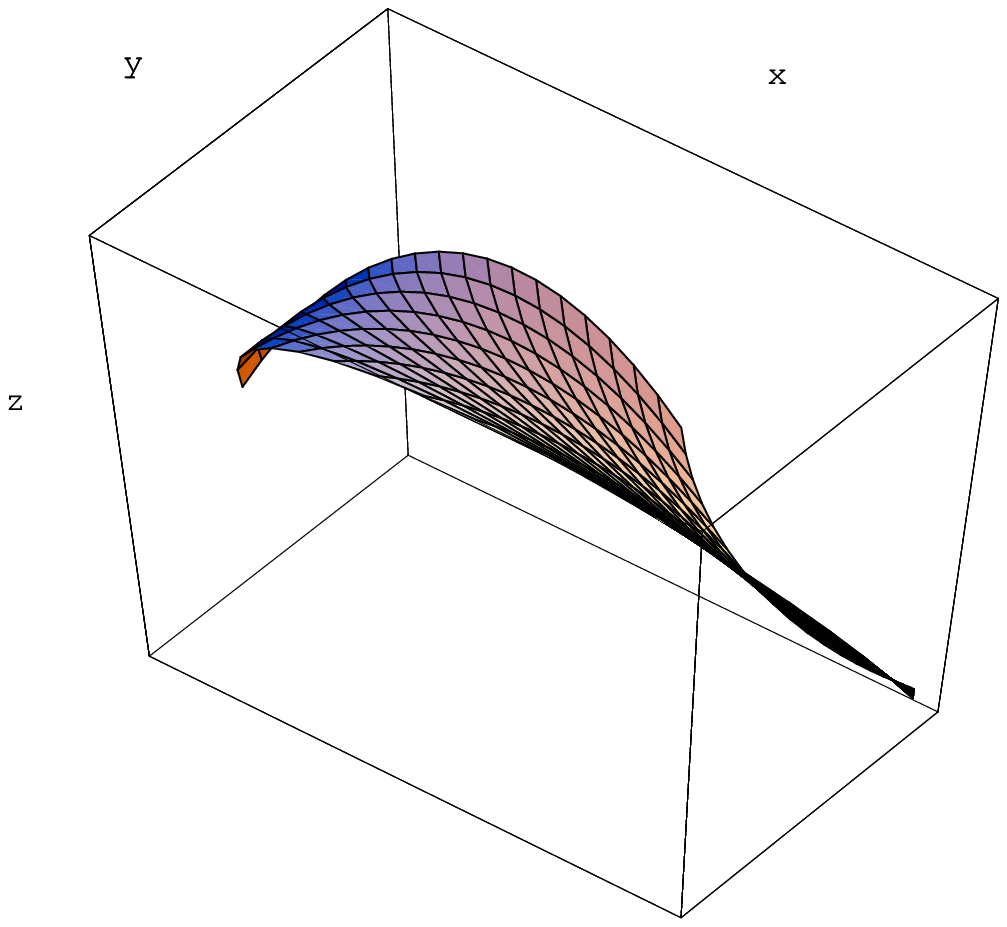}
\epsfxsize =5cm
\epsffile{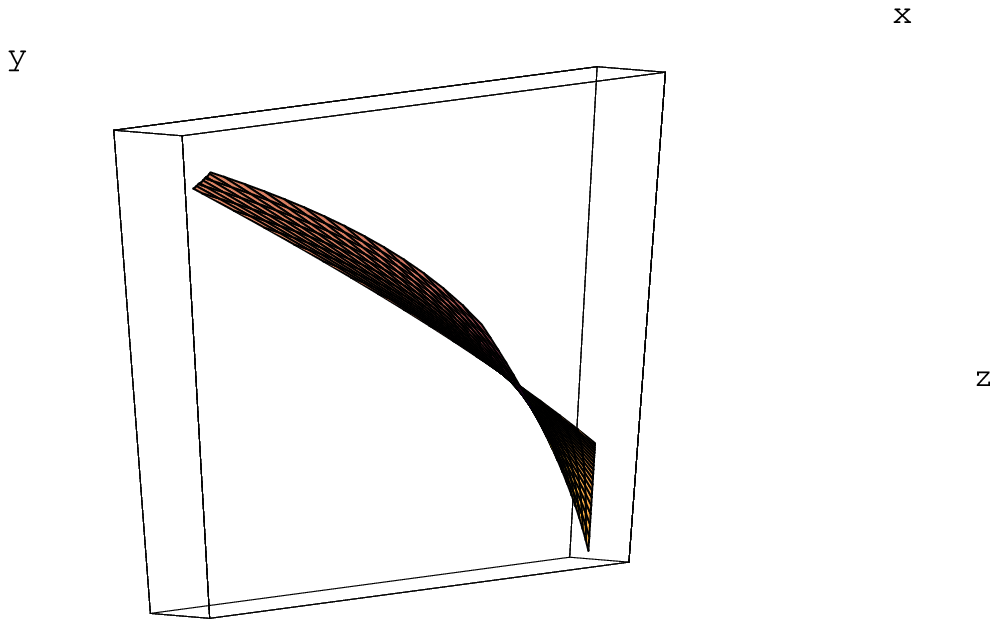}
\epsfxsize =5cm
\epsffile{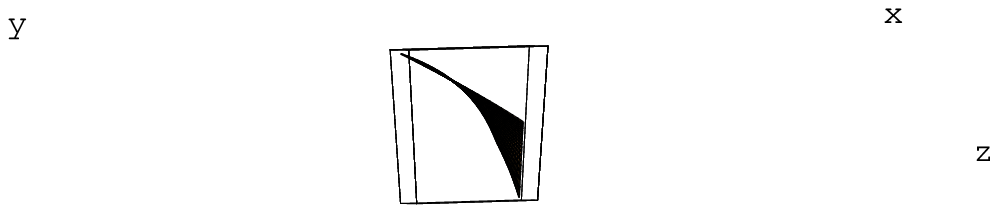}
\vskip 1cm
\caption{\label{congruomov}  In this  picture we present
the same congruence of geodesics at six different times of the
universe expansion, while $\Lambda$ grows and $\Delta$, after reaching
a maximum, decreases.
 }
}
\hskip 2cm
\unitlength=1.1mm
\end{center}
\end{figure}
\fi
\par
Having illustrated the shape and the properties of the geodesics for the three
dimensional sections of space--time we can now address the question
of geodesics for the full space--time. To this effect we calculate
first the three dimensional line element along the geodesics and we
obtain the following result
\begin{eqnarray}
&&\frac{d\ell^2\left(t, \lambda\right)}{d\lambda^2}  \equiv \Lambda(t) \, \left[  \,
 \dot{x}^2(\lambda) \, + \,  \, \dot{y}^2(\lambda) \right] +
 \Delta(t)\,   \left[ \dot{z}(\lambda)+ \frac{\varpi}{4} (x(\lambda) \, \dot{y}(\lambda)
 - y(\lambda) \, \dot{x}(\lambda)) \right]^2 = \nonumber\\
 && =  \frac{A_1^2}{96\Lambda(t)^2}\left( 16\,R^2\,\Lambda(t) \,{\varpi }^2 +
             \frac{{\left( -8\,\Lambda(t)  +
                            3\,\Delta(t) \,{\rho }^2\,{\varpi }^2 +
                            3\,R\,\Delta(t) \,\rho \,{\varpi }^2\,
                             \cos (\theta  - \varphi (\lambda )) \right) }
                      ^2}{\Delta(t) } \right) \equiv \nonumber\\
        &&\equiv F^2(t,\varphi) \, \left( \frac{d\varphi}{d\lambda}\right)^2
\label{delldlam}
\end{eqnarray}
In the last step of eq.(\ref{delldlam}) we have introduced the
notation:
\begin{equation}
F^2(t,\varphi) \, = \,  \left(R^2\,\Lambda(t) +
             \frac{{\left( -8\,\Lambda(t)  +
                            3\,\Delta(t) \,{\rho }^2\,{\varpi }^2 +
                            3\,R\,\Delta(t) \,\rho \,{\varpi }^2\,
                             \cos (\theta  - \varphi (\lambda )) \right) }
                      ^2}{16 \,\varpi^2 \Delta(t) } \right)
\label{F2tfi}
\end{equation}
and we have used  relation (\ref{dfidlam}).
\par
Hence we obtain the complete space--time geodesics  from those of three--space by
solving the following equation that relates the time coordinate $t$
to the angular coordinate $\varphi$:
\begin{equation}
-A(t) \, \left(  \frac{dt}{d\varphi}\right)^2 + F^2(t,\varphi) = k \,
\frac{4}{\varpi^2}\, \frac{A_1}{\Lambda^2(t)} \quad ; \quad \left \{
\begin{array}{cccl}
  k & = & -1 & \mbox{time--like}  \\
  k & = & 0 & \mbox{null--like}\\
 k & = & 1 & \mbox{space--like}
\end{array} \right.
\label{giuncata}
\end{equation}
Furthermore, the constant $A_1$ is inessential and can always be
fixed to $1$ since it can be traded for the constant $A_2$ which does
not appear in the equation. The differential eq.(\ref{giuncata})
appears rather involved since $F^2(t,\varphi)$ depends both on time and
the angle $\varphi$. Yet we can take advantage of the homogeneous
character of our space--time and simplify the problem very much.
Indeed due to homogeneity it suffices to consider the geodesics whose
projection in the $xy$ plane is a circle centered at the origin and
of radius $R$. All other geodesics with center in some point $\{ x_0
\, , \, y_0 \}$ can be obtained from these ones by a suitable
isometry that takes $\{0,0\}$ into $\{ x_0
\, , \, y_0 \}$. So let us consider geodesics centered at the origin
of the $xy$ plane. This corresponds to setting $\rho =0$. In this
case we obtain:
\begin{equation}
  F^2(t,\varphi)|_{\rho=0} \equiv F^2_0(t) \, = \, \frac{\Lambda(t) \,\left( 4\,\Lambda(t)  +
             R^2\,\Delta(t) \,{\varpi }^2 \right) }{\Delta(t) \,
        {\varpi }^2}
\label{pergolesi1}
\end{equation}
which depends only on time and the geodesic equations are reduced to
quadratures since we get:
\begin{equation}
  \int_0^{\varphi_{max}} \, d\varphi = \int_{-\infty}^{t_0} \frac{\sqrt{A(t)}}{\sqrt{F^2(t) - \frac{4\,
  k}{\varpi^2 \, \Lambda^2(t)}}} \, dt
\label{integralus}
\end{equation}
The convergence or divergence of the second integral in
eq.(\ref{integralus}) determines whether or not there are particle
horizons in the considered cosmology. Curiously, such horizons appear
as an angular deficit. For each chosen radius $R$ one can explore the
geodesic (which is a spiral) only up to some maximal angle
$\varphi_{max}$ at each chosen instant of time.
\subsection{Summarizing}
Summarizing the above discussion we can say that each exact
solution for a Bianchi type 2A cosmology, with or without matter,
presents a typical feature which we can generically name \textit{a
billiard feature}. In lack of isotropy, the scale factors
associated with the different dimensions (in this case we do not
have cartesian dimensions, yet we can identify the notion of
dimensions with the generators of the translation isometry
algebra) undergo a quite different fate. Two dimensions grow
indefinitely as in a isotropic big bang model, while the third
expands to a maximum, then it contracts and tends to zero. The
parameter governing this bending is $\varpi$, namely the only non
vanishing structure constant that deforms the Heisenberg algebra
away from an abelian algebra. An indication that this behaviour is
related to branes is evident in the example of the space
$0$--brane solution. There we observe that the direction which
undergoes the billiard phenomenon is the direction in which lies
the vector field, namely the $1$--form $A$, while those which
expand indefinitely are the transverse ones. This is exactly the
same as it was observed, for higher dimensions in paper
\cite{piervoiastatia}. There it was shown  how the
$\mathrm{A_2}$ solutions of the $E_{8(8)}/SO(16)$ sigma model
could, in particular, be oxided to $D=10$ supergravity backgrounds
containing $D3$--space branes. The dimensions of the brane
underwent a maximum and then decayed to zero, while for the
transverse ones the opposite was true. They were depressed at the
moment the brane dimensions were enlarged and then expanded again
while the parallel ones contracted. In this section we have
examined the geometric and physical implications of this peculiar
behaviour of the scale factors and we have explored the structure
of the canonical metric representative of $A_2$ models. From
the analysis of the previous sections, we know that these peculiar
Heisenberg algebra cosmologies are dual to any other $A_2$
solution, since there is just one Weyl orbit of
$\mathrm{A_2}$ embeddings.
\section{Conclusions and Perspectives}
\label{conclu} In the present paper we have explored the structure
of Weyl orbits for the embedding of regular subalgebras
$\mathbf{G}_r \hookrightarrow \mathrm{E_{8(8)}}$. The relevance of
this algebraic construction is that regular subalgebras of
$\mathrm{E_{8(8)}}$ generate exact time dependent solutions of the
sigma model $\mathrm{E_{8(8)}/SO(16)}$ and their embeddings
determine the oxidation of such solutions to exact time dependent
solutions of supergravity  in ten dimensions. In particular we
have considered the Weyl orbits of $\mathrm{A_r}$ subalgebras and
we have shown that there is only one orbit up to $r=6$. In a
future publication we plan to study the embeddings of other chains
of subalgebras, for instance the $\mathrm{D_r}$ chain. The
algebraic setup has been completely fixed here. For the
$\mathrm{A_r}$ chain we have shown that in the unique Weyl orbit
there is always a canonical representative that corresponds to a
pure metric configuration in dimension $d=3+r$. For the
$\mathrm{A_2}$ case the canonical metric representative is
related to Bianchi type 2A homogeneous cosmologies based on the
Heisenberg algebra. Through this relation we were able to present
some new exact solutions of matter coupled Einstein theory in this
Bianchi class that, up to our knowledge, were so far undiscovered.
We made an extensive analysis of their geometrical properties and
of their behaviour.
\par
As we already pointed out in the previous paper
\cite{piervoiastatia}, there are three main directions to be explored
in connection with the present new developments.
The first is the extension of our analysis to affine and hyperbolic
algebras. This means first reduce to dimensions $D=1+1$ or $D=1+0$
and then oxide back to $D=10$. In this process new classes of solutions
can be discovered, that include and extend the Geroch group. The
second line of investigation is the application of our algebraic
technique of deriving solutions to other low parameter cases, for
instance the dependence on light--like coordinates, leading to the
classification of gravitational waves. The third line of
investigation is the microscopic interpretation of these classical supergravity
solutions in terms of time--dependent boundary states and
space--branes. To this effect, as we explained in the introduction,
a firm control on the structure of Weyl orbits is particularly vital.
Indeed it allows to duality rotate classical solutions to others that
have a clear $D$--brane description.
\par
We plan to address all these questions in next coming future
publications.


\begin{thebibliography}{99}
\bibitem{Kachru:2003sx}
S.~Kachru, R.~Kallosh, A.~Linde, J.~Maldacena, L.~McAllister and
S.~P.~Trivedi, \emph{Towards inflation in string theory},
[arXiv:hep-th/0308055].
\bibitem{Kachru:2003aw}
S.~Kachru, R.~Kallosh, A.~Linde and S.~P.~Trivedi, \emph{De Sitter
vacua in string theory} Phys.\ Rev.\ D {\bf 68} (2003) 046005
[arXiv:hep-th/0301240].
\bibitem{Fre:2002pd}
P.~Fre, M.~Trigiante and A.~Van Proeyen, \emph{Stable de Sitter
vacua from N = 2 supergravity}, Class.\ Quant.\ Grav.\  {\bf 19}
(2002) 4167 [arXiv:hep-th/0205119];
M.~de Roo, D.~B.~Westra, S.~Panda and M.~Trigiante,
\emph{Potential and mass-matrix in gauged N = 4 supergravity},
JHEP {\bf 0311} (2003) 022 [arXiv:hep-th/0310187].
\bibitem{Gutperle:2002ai}
M.~Gutperle and A.~Strominger, \emph{Spacelike branes}, JHEP {\bf
0204} (2002) 018 [arXiv:hep-th/0202210].
\bibitem{iva}V.~D.~Ivashchuk and V.~N.~Melnikov, \emph{Multidimensional
classical and quantum cosmology with intersecting  p-branes}, J.\
Math.\ Phys.\  {\bf 39} (1998) 2866 [arXiv:hep-th/9708157];
\bibitem{cornalba}L.~Cornalba, M.~S.~Costa and C.~Kounnas, \emph{A resolution of the
cosmological singularity with orientifolds}, Nucl.\ Phys.\ B {\bf
637} (2002) 378 [arXiv:hep-th/0204261];
L.~Cornalba and M.~S.~Costa, \emph{On the classical stability of
orientifold cosmologies}, Class.\ Quant.\ Grav.\  {\bf 20} (2003)
3969 [arXiv:hep-th/0302137].
\bibitem{Papadopoulos:2002bg}
G.~Papadopoulos, J.~G.~Russo and A.~A.~Tseytlin, \emph{Solvable
model of strings in a time-dependent plane-wave background},
Class.\ Quant.\ Grav.\  {\bf 20} (2003) 969
[arXiv:hep-th/0211289].
\bibitem{que} F.~Quevedo, \emph{Lectures on string / brane
cosmology}, [arXiv:hep-th/0210292].
\bibitem{GV} M.~Gasperini and G.~Veneziano, \emph{The pre-big bang scenario
in string cosmology}, [arXiv:hep-th/0207130].
\bibitem{craps} B.~Craps, D.~Kutasov and G.~Rajesh, \emph{String
propagation in the presence of cosmological singularities}, JHEP
{\bf 0206}, 053 (2002) [arXiv:hep-th/0205101].
\bibitem{ban} T.~Banks and W.~Fischler, \emph{M-theory observables for
cosmological space-times}, [arXiv:hep-th/0102077].
\bibitem{setu} J.~Khoury, B.~A.~Ovrut, N.~Seiberg, P.~J.~Steinhardt and
N.~Turok, \emph{From big crunch to big bang}, Phys.\ Rev.\ D {\bf
65}, 086007 (2002) [arXiv:hep-th/0108187].
\bibitem{cope}
J.~E.~Lidsey, D.~Wands and E.~J.~Copeland, \emph{Superstring
cosmology}, Phys.\ Rept.\  {\bf 337}, 343 (2000)
[arXiv:hep-th/9909061].
\bibitem{mart} A.~E.~Lawrence and E.~J.~Martinec, \emph{String field theory
in curved spacetime and the resolution of spacelike
singularities}, Class.\ Quant.\ Grav.\  {\bf 13}, 63 (1996)
[arXiv:hep-th/9509149].
%
\bibitem{Sen:2002vv}
A.~Sen, \emph{Time evolution in open string theory} JHEP {\bf
0210} (2002) 003 [arXiv:hep-th/0207105].
\bibitem{Sen:2002nu}
A.~Sen,  \emph{Rolling tachyon}, JHEP {\bf 0204} (2002) 048
[arXiv:hep-th/0203211].
\bibitem{experiment} Riess A G {\it et al.} 1998 {\it Astron.\ J.}
 {\bf 116}1009, Perlmutter S {\it et al.} 1999 {\it Astron.\ J.} {\bf 517} 565,
 Sievers J L  {\it et al.} 2002 {\it Preprint}
 [arXiv:astro-ph/0205387]
\bibitem{linde90} Linde A D 1990 \emph{ Particle Physics and Inflationary Cosmology}
 (Switzerland: Harwood Academic)
\bibitem{piervoiastatia} P.~Fre, V.~Gili, F.~Gargiulo, A.~Sorin, K.~Rulik and M.~Trigiante,
\emph{Cosmological backgrounds of superstring theory and solvable
algebras: Oxidation and branes}, [arXiv:hep-th/0309237].
\bibitem{bill99}V.~D.~Ivashchuk and V.~N.~Melnikov,\emph{
Billiard representation for multidimensional cosmology with
intersecting p-branes near the singularity}, J.\ Math.\ Phys.\
{\bf 41} (2000) 6341 [arXiv:hep-th/9904077].
\bibitem{dualiza2} T.~Damour, M.~Henneaux and H.~Nicolai,
\emph{Cosmological billiards},
Class.\ Quant.\ Grav.\  {\bf 20} (2003) R145
[arXiv:hep-th/0212256].
\bibitem{Henneaux:2003kk}
M.~Henneaux and B.~Julia,
\emph{Hyperbolic billiards of pure D = 4 supergravities},
JHEP {\bf 0305} (2003) 047
[arXiv:hep-th/0304233].
\bibitem{deBuyl:2003za}
S.~de Buyl, G.~Pinardi and C.~Schomblond, \emph{Einstein billiards
and spatially homogeneous cosmological models},
[arXiv:hep-th/0306280].
\bibitem{Damour:2002tc}
T.~Damour, M.~Henneaux, A.~D.~Rendall and M.~Weaver,
\emph{Kasner-like behaviour for subcritical Einstein-matter systems},
Annales Henri Poincare {\bf 3} (2002) 1049
[arXiv:gr-qc/0202069].
\bibitem{Damour:pq}
T.~Damour and M.~Henneaux,
\emph{Chaos In Superstring Cosmology},
Gen.\ Rel.\ Grav.\  {\bf 32} (2000) 2339.
\bibitem{Damour:2001sa}
T.~Damour, M.~Henneaux, B.~Julia and H.~Nicolai,
\emph{Hyperbolic Kac-Moody algebras and chaos in Kaluza-Klein models},
Phys.\ Lett.\ B {\bf 509} (2001) 323
[arXiv:hep-th/0103094].
\bibitem{Damour:2000hv}
T.~Damour and M.~Henneaux,
\emph{E(10), BE(10) and arithmetical chaos in superstring cosmology},
Phys.\ Rev.\ Lett.\  {\bf 86} (2001) 4749
[arXiv:hep-th/0012172].
\bibitem{Damour:2000th}
T.~Damour and M.~Henneaux,
\emph{Oscillatory behaviour in homogeneous string cosmology models},
Phys.\ Lett.\ B {\bf 488} (2000) 108
[Erratum-ibid.\ B {\bf 491} (2000) 377]
[arXiv:hep-th/0006171].
\bibitem{Damour:2000wm}
T.~Damour and M.~Henneaux,
\emph{Chaos in superstring cosmology},
Phys.\ Rev.\ Lett.\  {\bf 85} (2000) 920
[arXiv:hep-th/0003139].
\bibitem{Demaret:sg}
J.~Demaret, Y.~De Rop and M.~Henneaux,
\emph{Chaos In Nondiagonal Spatially Homogeneous Cosmological Models In Space-Time Dimensions <= 10},
Phys.\ Lett.\ B {\bf 211} (1988) 37.
%
\bibitem{bertolotrigiaseries}
M.~Bertolini, M.~Trigiante and P.~Fre, \emph{N = 8 BPS black holes
preserving 1/8 supersymmetry}, Class.\ Quant.\ Grav.\  {\bf 16}
(1999) 1519 [arXiv:hep-th/9811251];
M.~Bertolini and M.~Trigiante, \emph{Regular R-R and NS-NS BPS
black holes}, Int.\ J.\ Mod.\ Phys.\ A {\bf 15} (2000) 5017
[arXiv:hep-th/9910237];
M.~Bertolini and M.~Trigiante, \emph{Regular BPS black holes:
Macroscopic and microscopic description of the  generating
solution}, Nucl.\ Phys.\ B {\bf 582} (2000) 393
[arXiv:hep-th/0002191];
M.~Bertolini and M.~Trigiante, \emph{Microscopic entropy of the
most general four-dimensional BPS black  hole}, JHEP {\bf 0010}
(2000) 002 [arXiv:hep-th/0008201].
\bibitem{Bianchiorinal} L. Bianchi, \emph{Sugli spazi a tre dimensioni che ammettono un gruppo continuo
di movimenti}, Memorie di Matematica e di Fisica della Societ\'a
Italiana delle Scienze, Serie Terza {\bf 11} (1898) 267-352.
\bibitem{lupope} H.~L\"{u}, C.N.~Pope, K.~Stelle \emph{Weyl Group Invariance and p-brane multiplets},
Nucl.\ Phys.\ B {\bf 476} (1996) 89
[arXiv:hep-th/9602140].
\bibitem{solv}L.~Andrianopoli, R.~D'Auria, S.~Ferrara, P.~Fre and M.~Trigiante,
\emph{R-R scalars, U-duality and solvable Lie algebras}, Nucl.\
Phys.\ B {\bf 496} (1997) 617 [arXiv:hep-th/9611014];
L.~Andrianopoli, R.~D'Auria, S.~Ferrara, P.~Fre, R.~Minasian and
M.~Trigiante, \emph{Solvable Lie algebras in type IIA, type IIB
and M theories}, Nucl.\ Phys.\ B {\bf 493} (1997) 249
[arXiv:hep-th/9612202];
E.~Cremmer, B.~Julia, H.~Lu and C.~N.~Pope, \emph{Dualisation of
dualities. I}, Nucl.\ Phys.\ B {\bf 523} (1998) 73
[arXiv:hep-th/9710119].
\bibitem{bianchiGRliter} R.M. Wald, Phys. Rev. D{\bf 28} (1983) 2118;
A A. Coley and J. Wainwright Class. Quantum Grav. {\bf 9} (1992)
651; M.~Goliath and G.~F.~R.~Ellis, \emph{Homogeneous cosmologies
with cosmological constant}, Phys.\ Rev.\ D {\bf 60} (1999) 023502
[arXiv:gr-qc/9811068];
U.~S.~Nilsson and C.~Uggla, \emph{Stationary Bianchi type II
perfect fluid models}, [arXiv:gr-qc/9702039].
\bibitem{humphrey2} J. Humprey, \emph{Reflection groups and Coxeter
groups} Cambridge University Press 1990.
\end{thebibliography}
\end{document}